%% file: main.tex
\documentclass[screen,authorversion,nonacm]{acmart} %

\AtBeginDocument{%
  \providecommand\BibTeX{{%
    \normalfont B\kern-0.5em{\scshape i\kern-0.25em b}\kern-0.8em\TeX}}}

\setcopyright{acmcopyright}
\copyrightyear{2025}
\acmYear{2025}

\usepackage{soul}
\usepackage{tcolorbox}
\usepackage{xcolor}
\usepackage{comment}

\usepackage{pifont}  %
\usepackage{tikz}    %
\usepackage{xr}

\newcommand{\ulcolor}[2][Red]{\setulcolor{#1}\ul{#2}}

\begin{document}

\title{Towards User-Focused Research in Training Data Attribution for Human-Centered Explainable AI}
\author{Elisa Nguyen}
\email{elisa.nguyen@uni-tuebingen.de}
\orcid{0000-0003-0224-268X}
\affiliation{%
  \institution{T\"ubingen AI Center, University of T\"ubingen}
  \country{Germany}
}

\author{Johannes Bertram}
\orcid{0009-0004-8216-980X}
\affiliation{%
  \institution{T\"ubingen AI Center, University of T\"ubingen}
  \country{Germany}
}

\author{Evgenii Kortukov}
\orcid{0009-0008-2542-6119}
\affiliation{%
  \institution{Fraunhofer Heinrich Hertz Institute}
  \country{Germany}
}

\author{Jean Y. Song}
\orcid{0000-0003-4379-3971}
\affiliation{%
  \institution{Information and Interaction Design, Yonsei University}
  \country{South Korea}
}

\author{Seong Joon Oh}
\orcid{0000-0002-8985-7689}
\affiliation{%
  \institution{T\"ubingen AI Center, University of T\"ubingen}
  \country{Germany}
}

\renewcommand{\shortauthors}{Nguyen, et al.}

\begin{abstract}
    Explainable AI (XAI) aims to make AI systems more transparent, yet many practices emphasise mathematical rigour over practical user needs. We propose an alternative to this model-centric approach by following a design thinking process for the emerging XAI field of training data attribution (TDA), which risks repeating solutionist patterns seen in other subfields. However, because TDA is in its early stages, there is a valuable opportunity to shape its direction through user-centred practices. We engage directly with machine learning developers via a needfinding interview study (N=6) and a scenario-based interactive user study (N=31) to ground explanations in real workflows. Our exploration of the TDA design space reveals novel tasks for data-centric explanations useful to developers, such as grouping training samples behind specific model behaviours or identifying undersampled data. We invite the TDA, XAI, and HCI communities to engage with these tasks to strengthen their research's practical relevance and human impact.
\end{abstract}



\maketitle

\section{Introduction}
\label{sec:introduction}

\begin{figure}[tbp]
    \centering
    \includegraphics[width=0.7\linewidth]{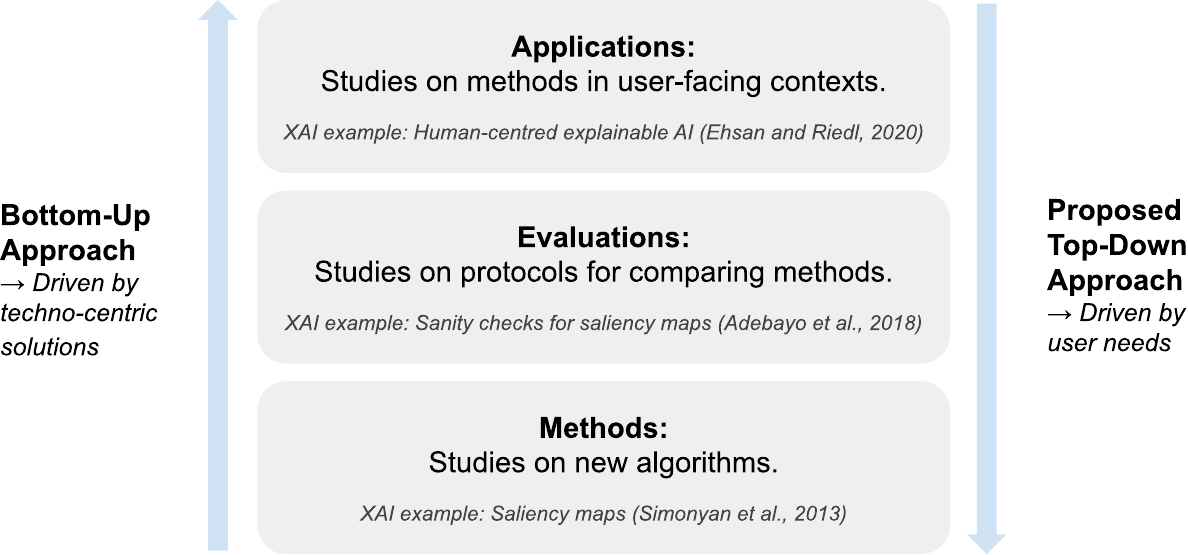}
    \caption{Comparison of bottom-up and proposed top-down development of research with selected examples from XAI: \cite{simonyan2013deep, adebayo2018sanity, ehsan2020hcxai}}
    \label{fig:research_approaches}
    \Description{Schematic showing the bottom-up and the proposed top-down research approaches. There are three boxes vertically stacked. The bottom box reads ``Methods: Studies on new algorithms. XAI example: Saliency maps (Simonyan et al., 2013)''. The middle box reads ``Evaluations: Studies on protocols for comparing methods. XAI example: Sanity checks for saliency maps (Adebayo et al., 2017)''. The top box reads ``Applications: Studies on methods in user-facing contexts. XAI example: Human-centred explainable AI (Ehsan and Riedl, 2020)''. There is an arrow pointing upwards to the left of the boxes, annotated with ``Bottom-Up Approach -> Driven by techno-centric solutions'' and an arrow pointing downwards to the right of the boxes, annotated with ``Proposed Top-Down Approach -> Driven by user needs''.}
\end{figure}

Explaining and understanding AI model behaviour is critical for the use and development of AI models in practice, especially in high-stakes domains such as medicine, law, or finance~\cite{aiact}.
However, despite significant progress in Explainable AI (XAI) techniques~\cite{guidotti2018, dwivedi2023explainable, nauta2023from}, the community has faced criticism for its predominant techno-centric focus on solutions as opposed to the sociotechnical problem of helping users understand model behaviour~\cite{ehsan2021expanding, williams2021towards, wolf2019explainability}.
We highlight that historically, XAI research has followed a bottom-up trajectory where methods and frameworks have built on each other: beginning with technical methods that are seemingly useful, then justifying practicality through quantitative evaluation frameworks, and ultimately extending to real-world applications through field studies and user studies (Figure~\ref{fig:research_approaches}).

An example of this bottom-up trajectory is the development of the subfield of feature attribution. Feature attribution explanations quantify each input feature's contribution to the model prediction, identifying the most influential inputs. 
One of the first feature attribution methods in the context of deep learning was presented in 2013, introducing the visualisation of neural network input gradients as saliency map explanations~\cite{simonyan2013deep}.
The paper provides 15 qualitative examples as evidence that saliency maps explain. 
Several other methods followed, either proposing changes to satisfy specific formal axioms (e.g.,~\cite{smilkov2017smoothgrad, sundararajan2017axiomatic}) or proposing more widely applicable model-agnostic methods (e.g.,~\cite{lundberg2017unified, ribeiro2016should}). 
Around 2017, XAI reached an inflection point where researchers began reassessing the practicality of explanation methods and introduced standardized evaluation protocols to systematically measure their effectiveness in achieving their explanatory goals~\cite{doshi2017towards, adebayo2018sanity, bilodeau2024impossibility}. Soon after, recommendations from the social science view of explanations emphasised that explanations are part of a social process, which shifted the field's focus to explanation effectiveness in real-world decision-making~\cite{miller2019explanation}. 
Increasing amount of works studying XAI started to focus on the users and XAI as part of a sociotechnical system~\cite{wang2019designing,ehsan2021expanding, brennen2020what, rutjes2019considerations, kim2022hive, kim2023help, fok2023search, kaur2020interpreting, mei2023users, ehsan2024who, lakkaraju2022rethinking}, 
some criticising the ``[solutionism (seeking technical solutions) and] formalism (seeking abstract, mathematical solutions)~\cite[p.1]{ehsan2021expanding}'' of early approaches. In 2020, this subfield was coined as human-centred XAI (HCXAI)~\cite{ehsan2020hcxai}.  

\begin{figure}[tp]
    \centering
    \includegraphics[width=\linewidth]{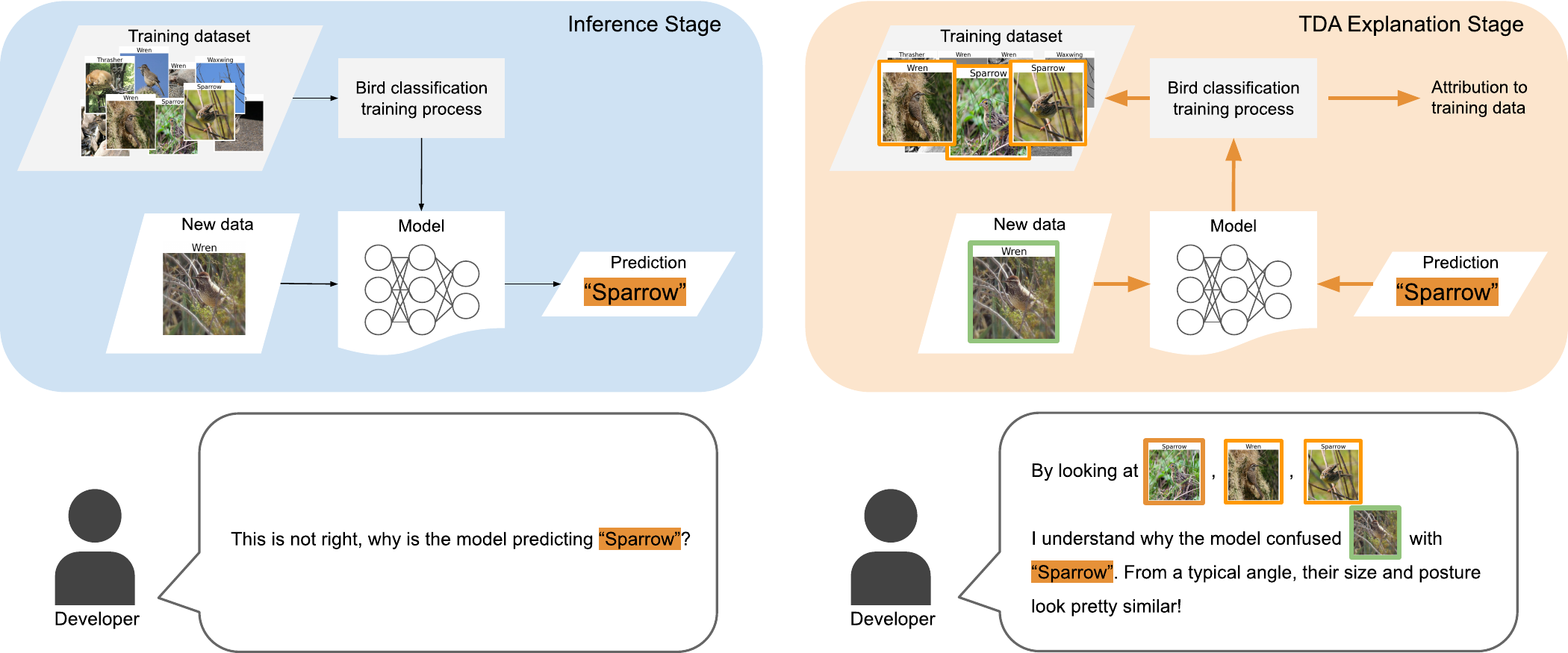}
    \caption{Example for the use of TDA as explanation in a fictional bird classification model development scenario. After training the classifier on the training data, the developer tests the model during the inference stage and inspects model errors (left). TDA explanations identify relevant training data to the misclassification, enabling the developer to build and refine hypotheses about reasons for the model error (right).}
    \label{fig:tda_example}
    \Description{This figure depicts a visualization of using TDA explanations for a bird classification model, consisting of two flowcharts next to each other. On the left, the flowchart shows the inference stage of a trained bird classifier: An input image is given to a neural network model, which outputs the wrong prediction ``Sparrow''. The flowchart shows that the model is the output of the training process, and the training data is input to the training process. Below this flowchart, a developer asks ``This is not right, why did the model prediction Sparrow?''. On the right side of the figure, the explanation generation stage through TDA is depicted. It is the same flowchart as on the left, except that arrows are pointing backward from the wrong prediction, through the model and training process to the training data. A few images in the training data are highlighted. Below the right flowchart, the developer says: ``By looking at the training data, I understand why the model confused the input with Sparrow. From a typical angle, their size and posture look pretty similar!''}
\end{figure}

Despite the efforts to rectify the solutionism and formalism in XAI communities, we observe repeating patterns in the relatively young field called training data attribution (TDA) explanations, where the model behaviour is explained based on the influence of the training data~\cite{Hammoudeh2022TrainingDI} (Figure~\ref{fig:tda_example}).
TDA was first introduced to understand deep learning models in 2017~\cite{kohliang2017}, and has produced several technical explanation approaches since then~\cite{charpiat_input_2021,guo-etal-2021-fastif, pruthi_estimating_2020, schioppa2022arnoldi, park2023trak, bae2024training, ilyas2025magic}. The field's development demonstrates a strong focus on seeking technical solutions similar to the trajectory of feature attribution research. 
We argue that TDA research is at a critical juncture, gaining momentum for its ability to address key issues such as data valuation~\cite{ghorbani2019data, choe2024your}, debiasing~\cite{jain2024data}, memorisation and copyright in foundational models~\cite{feldman2020neural, zheng2023intriguing, lin2024efficient, akyurek2022towards} and understanding model behaviour and errors~\cite{grosse2023studying, wang2024error}. Yet, it remains dominated by solutionism and formalism, with limited user-focused studies, potentially creating a rift between TDA methods and users' practical needs.

In this study, we bring users to the centre of our investigation on TDA as explanations, shifting the focus from a technology-centric perspective to a user-centric one. This approach enables researchers to develop practical solutions with better usability. We propose future research directions for the TDA and  HCI community that are entirely sourced and motivated by potential users of TDA. To guide a systematic exploration of user needs, we focus on users who traditionally have access and agency over the training data, model developers, as TDA explains model behaviour by identifying how each training sample is relevant to a model outcome. This study thus aims at contributing to the study of HCXAI~\cite{ehsan2020hcxai} and, more specifically, human-centred model development tools~\cite{kaur2020interpreting, cabrera2023what}. 

We break down our research question into three parts:
\begin{itemize}
    \item \textbf{RQ1:} What data-related explanation needs, both explicit and latent, do model developers have?
    \item \textbf{RQ2:} To what extent do existing TDA approaches align with model developers' needs?
    \item \textbf{RQ3:} What kind of information produced by TDA will be useful for model developers?
\end{itemize}

We conducted a two-stage needfinding study with model developers to address our research questions. 
Through a qualitative interview study (N=6), we explore potential usage scenarios of TDA explanations. We note that while TDA has the potential to support many data-centric tasks for model developers, it was not yet widely known or used in practice. 
Based on the interview findings, we derive a design space for TDA explanations representing different TDA tasks (e.g., identifying the training samples with the largest effect on model behaviour when removed in contrast to when given a larger weight during training). In a subsequent scenario-based interactive user study (N=31), we study which specific data-centric tasks are needed to support testing users' hypotheses. We identify tasks relevant to potential TDA users, but are largely overlooked by current TDA research. We invite the TDA and broader XAI research communities to address these gaps and reorient TDA research away from solutionism toward user-driven inquiry -- adopting a top-down, user-centred perspective. We advocate for adopting this approach in other areas of XAI and AI as a whole to avoid focusing on technically elegant solutions that may overlook user needs.

To summarise, our work makes the following contributions:
\begin{itemize}
    \item We propose a top-down approach to XAI research driven by user needs and demonstrate this approach on the emerging TDA subfield. 
    \item Through a two-stage needfinding study with model developers, we identified what data-related information developers need and how existing TDA approaches address the needs (\autoref{sec:method} and \autoref{sec:rq_answers}). 
    \item From the use cases we found, we identified research directions for user-focused TDA that are currently understudied and invite the community to study these (\autoref{sec:futurework}).
    \item We outline our approach into a high-level framework for needs-based research that connects user-oriented inquiry with method development in the early phases of technology research (\autoref{sec:needframework}). 
\end{itemize}

\section{Related Work}
\label{sec:rw}

Our work takes a user-centered approach to XAI research, building from empirical insights on training data attribution (TDA) use cases, which aligns it with human-centered explainable AI (HCXAI). 
To this end, we draw on needfinding methods, a user-centered research approach that helps uncover latent needs and usage scenarios, to inform the design and development of TDA tools. The following sections connect our work to existing literature in HCXAI and TDA.

\subsection{HCXAI for Model Development} 
The aim of explainable AI (XAI) research is to study methods that explain the decisions and behaviour of AI models commonly referred to as ``black boxes''~\cite{guidotti2018}. 
The field has developed various explanation methods over the years to explain different types of tasks and models.
Yet, AI explanations are only meaningful if they are interpretable by the intended users in a way that informs or supports their decision-making. This human aspect was often neglected in early XAI studies~\cite{rong2024hcxai, nauta2023from}. 
In the Human-centred XAI (HCXAI) field, explanations of black-box models are studied holistically as sociotechnical systems centred around the human who is interested in the explanation~\cite{ehsan2024who}.
HCXAI explores how to design AI technology and explanations that focus on user understanding and practicality~\cite{ehsan2020hcxai}. Work in this area covers multiple aspects that affect the user, e.g. human-AI-collaboration~\cite{kim2023help, shen2023convxai, taesiri2022visual, lakkaraju2022rethinking}, design guidelines~\cite{liao2020questionbank, hadash2022improving, wang2019designing}, evaluation of human understanding of explanations~\cite{rong2024hcxai, kim2022hive, hadash2022improving, rutjes2019considerations} and understanding user needs and requirements~\cite{juneja2024dissecting, kaur2020interpreting, brennen2020what, jin2023invisible}.

However, much XAI work has focused on system end-users who are affected by an AI system~\cite{juneja2024dissecting, jin2023invisible, kim2023help, shen2023convxai, taesiri2022visual, lakkaraju2022rethinking} and rather few explored the need for model developer who can modify the model architecture, parameters, or training process~\cite{kaur2020interpreting, hong2020human}.
Our work contributes to the HCXAI field by proposing an early consideration of user needs, especially the model developers' needs, to inform ongoing research in XAI methods. We focus particularly on TDA explanations for model developers because they have agency over the data, which have not been explored widely in HCXAI before. 

\subsection{TDA as Explanation}
Training data attribution (TDA) explains model predictions by pointing to training data samples relevant to the model predictions~\cite{Hammoudeh2022TrainingDI}. 
TDA differs from the mainstream XAI approach known as feature attribution (FA) where the model predictions are attributed to the features of an input given to the model, disregarding the impact of training data.
Interest in TDA has recently accelerated with the paradigm shift towards data-centric AI (DCAI)~\cite{DCAISurvey2024}, where the importance of using the right training data is emphasised over other factors like model architecture and optimisation algorithms~\cite{sambasivan2021cascades}. 
Under the context of DCAI, TDA has been used to enhance data quality by cleaning faulty data labels~\cite{teso2021interactive} and detecting model biases~\cite{pezeshkpour-etal-2022-combining, brunet2019understanding, wang2024error, jain2024data}, as well as answering questions around data valuation or memorisation and copyright issues in foundational models in recent years~\cite{choe2024your, zheng2023intriguing, grosse2023studying, lin2024efficient}. 
As such, TDA offers the potential of providing actionable explanations for users with dataset access, like model developers.
The recent rise of foundational models, such as large language models (LLMs) and diffusion-based image-generation models, further contributed to the interest in TDA, as it has proven effective for studying the inner mechanisms as well as understanding the associated copyright and privacy leakage issues therein~\cite{feldman2020neural,choe2024your, zheng2023intriguing, grosse2023studying, lin2024efficient}.
With the accelerated growth, we observe again a general focus on technical and mathematical solutions in the TDA community~\cite{kohliang2017,charpiat_input_2021, guo-etal-2021-fastif, schioppa2022arnoldi, pruthi_estimating_2020, park2023trak, bae2024training, choe2024your, kwon2023datainf, ilyas2022datamodels, wang2025better, wang2025capturing, ilyas2025magic}, without a deep understanding of user needs, even in studies applying TDA to a downstream tasks~\cite{feldman2020neural, jain2024data, zheng2023intriguing, akyurek2022towards, li2024delta}. Our work addresses this gap by studying practical needs and providing concrete recommendations for the field.

\section{Background: What is TDA?}
\label{sec:tda}

This section provides background on training data attribution (TDA), describing what it is and why it is an interesting explanation approach. 
In theory, TDA links specific model behaviour to the training data, treating the data as the root cause of learned behaviours~\cite{Hammoudeh2022TrainingDI}. TDA offers insights into the model by identifying the training data most influential to what the model has learned. 

\paragraph{Formal definition.} 
Let us consider a training sample $z_\mathrm{train}:=(x_{\mathrm{train}},y_\mathrm{train})$ and a test sample  $z_\mathrm{test}:=(x_{\mathrm{test}},y_\mathrm{test})$, where $x$ indicates the input and $y$ indicates the true answer the model is supposed to predict. 
An attribution method $\tau$ assigns a score to a training data point $z_\mathrm{train}$, 
as the change in the \textit{correctness} of the model prediction on $z_\mathrm{test}$, 
indicating its importance measured via the model loss $\mathcal{L}(f_\theta(x_{\mathrm{test}}),y_\mathrm{test})$, before and after removing the sample $z_\mathrm{train}$ from the training procedure~\cite{hampel1974}:
\begin{equation}
    \label{eq: loo}
    \tau(z_{\mathrm{train}}, z_{\mathrm{test}};f_\theta) := \mathcal{L}\left(f_{\theta_{\setminus z_{\mathrm{train}}}}(x_{\mathrm{test}}), y_\mathrm{test} \right) - \mathcal{L}\left(f_\theta(x_{\mathrm{test}}), y_\mathrm{test} \right).
\end{equation}
Here, we indicate the model trained with all training samples as $f_\theta$ and the one trained without $z_\mathrm{train}$ as $f_{\theta_{\setminus z_{\mathrm{train}}}}$.

\paragraph{Main focus of TDA research.}
Much of the research in the TDA community has centred on efficiently approximating Equation \ref{eq: loo} because computing $\tau$ directly by re-training a model without each training sample $z_\mathrm{train}$ is computationally prohibitive. To identify the most influential training sample, the re-training has to be repeated as often as the number of training samples. 
In 2017, Koh \& Liang \cite{kohliang2017} introduced a gradient-based approximation called influence functions (IF) for deep models. Since IF was still computationally prohibitive due to the need to compute and invert a massive Hessian matrix, subsequent works in 2020 and 2021 have focused on speeding up the algorithm by trading off precision and computational costs \cite{pruthi_estimating_2020, charpiat_input_2021, schioppa2022arnoldi, guo-etal-2021-fastif}.
After the advent of ChatGPT in late 2022 \cite{OpenAI_ChatGPT35} and the boom in large language models in 2023 \cite{naveed2023comprehensive}, the community has developed further enhancements to apply the TDA methods on billion-scale LLMs
\cite{bae2024training, choe2024your, grosse2023studying, park2023trak}. For a survey on TDA methods, we refer the reader to \citet{Hammoudeh2022TrainingDI}.

\paragraph{Critique.}
We observe hints of solutionism and formalism in the development of TDA research, as the majority of the community effort is dedicated to addressing computational inefficiencies of the method itself, without questioning the practical relevance and user values of TDA as defined in Equation~\ref{eq: loo}.
This particular formulation of TDA has not gone without criticism within the community: a prior work~\cite{nguyen2023bayesian} discussed the statistical insignificance of removing a single training sample in modern neural network training which affects the fragility of approximation methods~\cite{basu_influence_2021, bae_if_2022} and \cite{ilyas2022datamodels} recognised the limitation and argued for examining the removal of multiple training samples.
Some TDA works have explored possible applications that are potentially interesting to developers such as fixing mislabelled data~\cite{kohliang2017} or identifying brittle classes~\cite{ilyas2022datamodels}. 
However, these applications are not \emph{driven} by user needs and rather targeted toward the evaluation of TDA methods, potentially missing application scenarios and not addressing user needs. 

\paragraph{Potential.}
TDA provides data-centric information that users can make use of to gain a deeper understanding of their training datasets. Such information may be particularly interesting if the user has agency over the training dataset, meaning that they are in a position to adjust it. Hence, TDA can be especially useful to model developers when a high-quality training dataset is of crucial importance~\cite{sambasivan2021cascades}. Developers often make sense of their models through manually inspecting training data~\cite{cabrera2023what}. There are tools that enable model developers to explore datasets by offering a visual interface~\cite{hohman2020understanding, piorkowski2023aimee}, but they often do not inform about the connection between model output and training data. As a data-centric approach to gaining such insight, TDA could support and enhance model developers' data-centric work. Currently, there is no known work in the TDA domain involving actual users to identify their needs to assess the relevance of the current TDA tasks and identify research directions grounded in user needs that the community should focus on.

\section{Needfinding: Asking Model Developers What They Really Need}

\label{sec:method}

Our work investigates the user needs surrounding training data attribution (TDA) explanations to support a user-centered development for machine learning (ML) developers and practitioners. To this end, we conducted a two-stage needfinding study: first, we ran an interview study with ML practitioners (N=6) to identify the design space of TDA explanations that represent useful data-centric information for model development. 
Second, we build on this space through a scenario-based interactive user study (N=31) aimed at eliciting concrete user needs and expectations. 
IRB approval was obtained from one of the authors' institutions. 

\subsection{Interview Study Procedure}
\label{method:interview}

We conducted semi-structured interviews with machine learning (ML) developers who work outside academic research.
The goal of the interview was to explore developers' current and anticipated practices around training data to surface latent needs that may inform the design of TDA explanation tools. 

\subsubsection{Participants}
\label{interview_participants} 
To get a realistic impression of user needs for TDA explanations in practice, we target participants who develop ML applications outside of academia. 
Our inclusion criteria were: Participants should (1) have at least one year of experience in developing ML systems and (2) work in a high-risk application area according to the EU AI Act~\cite{aiact} (e.g., health care, cars, law enforcement. Complete list in Appendix~\ref{appendix:high-riskapps}). 
The first criterion ensures that participants have hands-on experience in developing machine learning models which their statements can be rooted in. The latter criterion serves to find participants who are likely to use explanations, as these areas are subject to further regulations~\cite{guidotti2018, doshi2017towards}.
Recruiting participants poses a challenge, especially in high-risk application areas. Hence, we used purposive sampling~\cite{guest_collecting_2023} and approached participants from the authors' network \footnote{As the interview study is of exploratory nature, we stopped recruitment for the interviews after three months. We note that potential participants in legal and finance domains were unable to take part in the study due to non-disclosure constraints.}.
We recruited six participants from various domains and degrees of experience as summarised in Table~\ref{tab:participants_interview}.

\begin{table}[t]
    \centering
    \caption{Interview participant information.}
    \small
    \begin{tabular}{@{}llll@{}}
        \toprule
        \textbf{ID}  & \textbf{Location} & \textbf{Domain} & \textbf{Job experience / with ML} \\ \midrule
        I1               & US             & Autonomous vehicles  & 2 years / 7 years                 \\
        I2               & NL     & Telecommunications  & 3 years / 5 years                   \\
        I3                & FI         & Computer vision for automation  &    4 years / 6 years                         \\
        I4               & NL     & Health     & 1 year / 3 years                   \\
        I5             & BE         & Health      & 2 years / 6 years                 \\
        I6              & PK        & Health       &   5 years / 2 years                          \\
        \bottomrule
    \end{tabular}
    \label{tab:participants_interview}
\end{table}

\subsubsection{Interview process}
The interviews were conducted in person or remotely through video call using Zoom\footnote{\url{https://zoom.us/}}.
All interviews were one-on-one conversations. The participants were first informed about the purpose of the study and data processing. Upon receiving informed consent, we began the interview recording. 
The interviews lasted between 30 and 60 minutes.  
Each interview addressed the following topics:
\begin{itemize}
    \item \textbf{Job Roles and Responsibilities.} Perspectives may vary between different domains, levels of seniority, and experience. This information provides context about the interview responses for subsequent analysis.
    \item \textbf{ML Model Development Workflow.} By asking about interviewees' development workflows, we aim to uncover diverse development styles and the challenges they face, both explicit and latent. This helps surface underlying needs that may not be readily articulated, but could inform opportunities for supporting these workflows through TDA tools.
    \item \textbf{Uses and Relevance of Training Data.} We explicitly asked participants about the role that training data plays in their day-to-day tasks. Their responses highlight the types of data-centric information they find most relevant, and the specific tasks or decision points where such information becomes critical.
    \item \textbf{Perspectives on XAI and TDA.} 
    We address the participant's perspectives on XAI and particularly on TDA to understand current explanation usage scenarios in model development and explore participant's opinions about TDA, a data-centric approach to XAI.
\end{itemize}

\subsubsection{Data analysis}
The interviews were transcribed using Whisper~\cite{whisper} and translated using DeepL\footnote{\url{https://www.deepl.com/translator}} if necessary. Then they were manually cleaned up and pseudonymised by the authors. 
The transcripts were analyzed through an inductive thematic analysis~\cite{braun2012thematic} by two coders, where the codes were directly annotated in the transcript. The analysis was iterative: One interview transcript from a pilot interview was first jointly analysed in an initial coding workshop that resulted in an initial set of themes. Afterwards, the coders independently code three transcripts at a time, expanding on the themes found in the initial analysis.
During an intermediate coding workshop, agreements and disagreements between the coder's themes were discussed. The workshop resulted in a merged definition of the themes that were used for the remaining transcripts. At the intermediate coding workshop, the inter-rater agreement measured by Krippendorff's $\alpha$~\cite{krippendorff2011computing} was $\alpha=0.832$ computed with \cite{marzi2024k}.
The final coding workshop took place after both coders have reviewed the remaining transcripts. The final inter-rater agreement was $\alpha=0.846$.

\subsection{Interview Study Findings}
\label{results:interviews}

\definecolor{ourorange}{RGB}{246,178,107}
\setul{-.2em}{0.4em}

\begin{figure}
    \centering
    \includegraphics[width=0.9\linewidth]{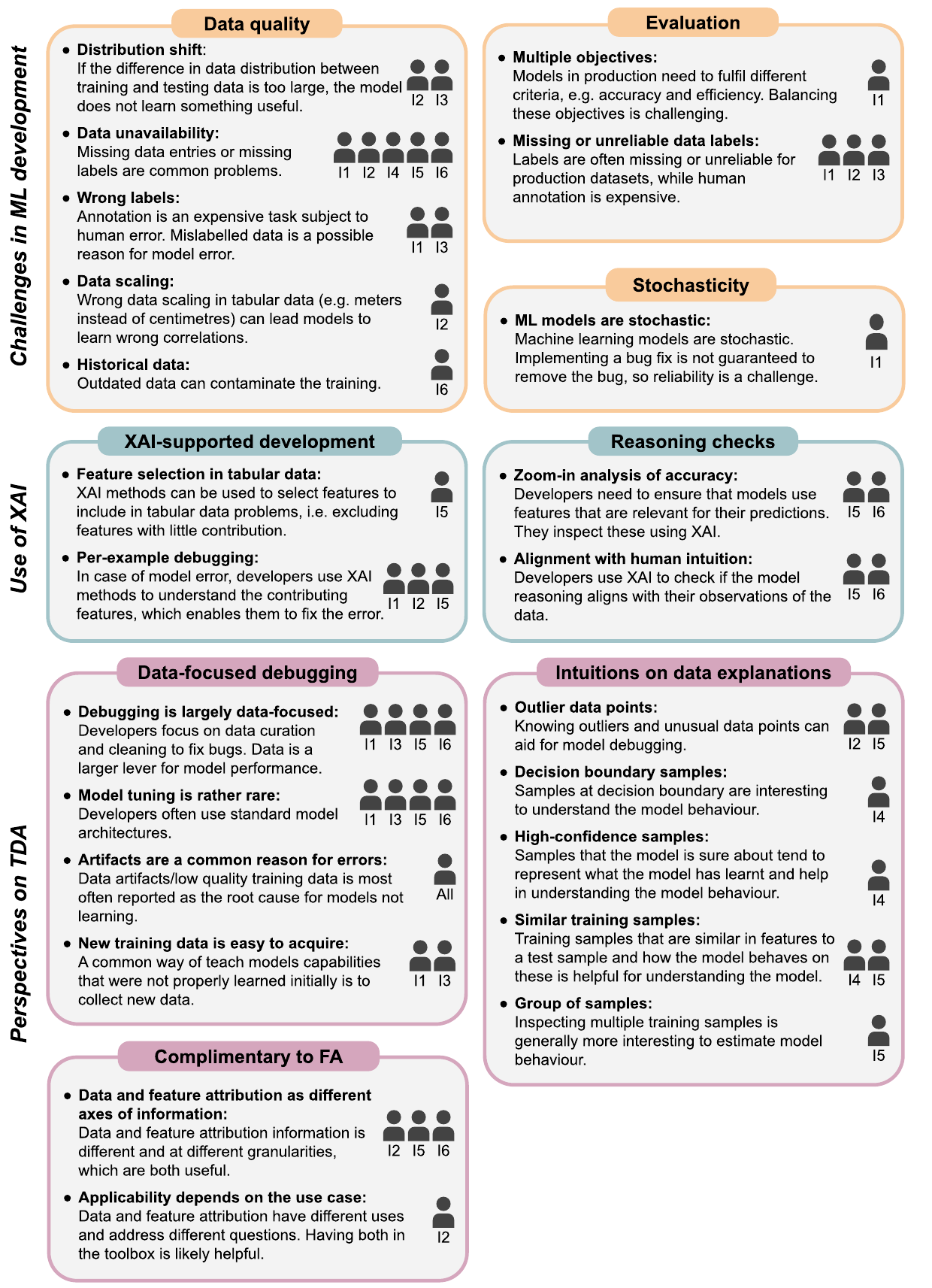}
    \caption{Codes and themes on developers' current and anticipated practices on training data.}
    \label{fig:interview_tdacodes}
    \Description{The figure shows the outcomes of the thematic analysis, eight themes altogether, grouped into three meta-themes: Challenges in ML development (Data quality, Evaluation, Stochasticity), Use of XAI (XAI-supported development, Reasoning checks) and Perspectives on TDA (Data-focused debugging, Complimentary to FA, Intuitions on data explanations).}
\end{figure}

The analysis resulted in themes within three key meta-themes: the main challenges in working with ML systems, whether and how explanation tools are used, and perspectives and opinions about data-centric explanations like TDA. The meta-themes encompass themes and are depicted in Figure~\ref{fig:interview_tdacodes}. We elaborate on each meta-theme below:

\subsubsection{Challenges in ML development}
In the interviews, we asked participants about challenges in their development workflow to explore whether XAI can address them. We identified three TDA-related themes under this meta-theme with several challenges: Data quality issues are often the root cause of model malfunction (Theme: data quality) (I1, I3, I4, I5, I6): \textit{``the models can only be as good as the [...] data that you feed in."} (I3). For instance, participants encountered difficulties developing models due to missing data or absent labels (I1, I3, I4, I5, I6). Our interviews support previous observations that a significant part of model development is data work~\cite{sambasivan2021cascades, DCAISurvey2024}. 
Besides data quality, our interviews revealed that the evaluation and quantification of ML systems' performance poses a challenge in the development pipeline as well (Theme: Evaluation). In practice, developers need to balance different desiderata and objectives from their businesses and stakeholders. For example, the priority of predictive accuracy versus efficiency varies according to the particular business scenario and application. However, a large challenge in evaluation is also linked to data quality since evaluation datasets in practice are changing across time and tasks, so that missing data and unreliable data labels are a common issue.
Also, the stochastic nature of ML models (I1) are pain points for the development and evaluation of ML systems since bug fixes may not work in the way they were intended, requiring further analysis (Theme: Stochasticity). Due to this stochasticity, participants perceive additional data collection as a more reliable bug fix: \textit{``[If] you have a neural network system, if you think you've solved the problem, you're never 100\% guaranteed [...] In practice, [...] whenever we had an issue [...] we would first like get a lot more data, label a lot more data and add those new examples to the training data, train a new model on these.'' (I1)}.

\subsubsection{Use of (feature attribution) XAI}
This meta-theme groups two themes about the use or lack of XAI in practice. Our analysis revealed that, in most cases, when developers claimed to have used XAI, they were referring to feature attribution-based techniques (I1, I3, I5) (Theme: XAI-supported development). We found that feature attribution-based XAI offers explanations for per-example debugging or acts as a sanity check for model reasoning (I5, I6). Participants also use XAI to understand real-world phenomena modelled by their ML system: \textit{``The valuable part is that this model that can predict [...] can also show us why [...] when we use some explainability techniques."} (I2) and as a communication means with their business counterparts or to get customer buy-in for their ML systems: \textit{``[We] output [the explanation] to the business basically. That's what they work with."} (I2), \textit{``[It's] all about the buy-in that you get, and SHAP definitely helps."} (I5) (Theme: Reasoning checks). 
While explanations have different purposes, we note that participants use XAI tools mainly as an out-of-the-box functionality (e.g., the SHAP library~\cite{lundberg2017unified}). We find that implementation thresholds must be low for the adoption of XAI in practice. 

\subsubsection{Perspectives on TDA} 
None of the participants were familiar with TDA, highlighting a gap between research and practice and emphasising that TDA technology is in its early stages. Our analysis, based on discussions about the concept of TDA, identifies three themes reflecting participants' views on XAI and TDA. Overall, participants find XAI useful for debugging and communication, though they agree its usefulness is use-case-dependent and not always guaranteed. They expect TDA not to be an exception (I2, I5) but are optimistic about its potential, especially as data is seen as the most important variable in model development (I1, I2, I3, I5): \textit{``[What] drives your model is your data. [...]''} (I5) (Theme: Data-focused debugging). In line with this theme, all participants also agreed that the model debugging process is largely data-focused, with training data being viewed as a lever for model performance: \textit{``It's often the most, it's often the biggest lever that we have in solving a problem."} (I1) (Theme: Data-focused debugging). Participants were curious whether TDA could be a potential tool for understanding bugs and enabling them to curate their datasets effectively. When comparing TDA to other XAI approaches like feature attribution (FA), participants point out that the TDA and FA target insights about model behaviour at different granularities (I2, I5, I6) and could be complimentary (Theme: Complimentary to FA). For instance, I1 mentioned that TDA could save time by identifying faulty training data while I2 pointed out that TDA and FA could be used together to identify the relationship between the faulty training data and learned features. We observed that participants have different notions of interesting training samples, with some focused on finding faulty data (I1) and others on identifying samples similar to a specific one (I5, I6) (Theme: Intuitions on data explanations). \\

In conclusion, our interview study highlights that TDA has significant potential to address persistent data-centric challenges in model development, particularly in \emph{handling training data quality} and \emph{data-focused debugging}. However, participants revealed divergent mental models of what TDA entails and what they consider as a relevant sample to identify. This suggests a conceptual gap that may hinder effective adoption of TDA-based data-centric explanations. 
This motivated us to further investigate which specific forms of data-centric information practitioners actually need and value during model development. In the second stage of our needfinding study, we explore these information needs to better align TDA-driven tools with real-world development workflows.

\subsection{Defining a design space of TDA explanations}
\label{sec:design_space}

\begin{figure}[t]
    \centering
    \includegraphics[width=0.95\linewidth]{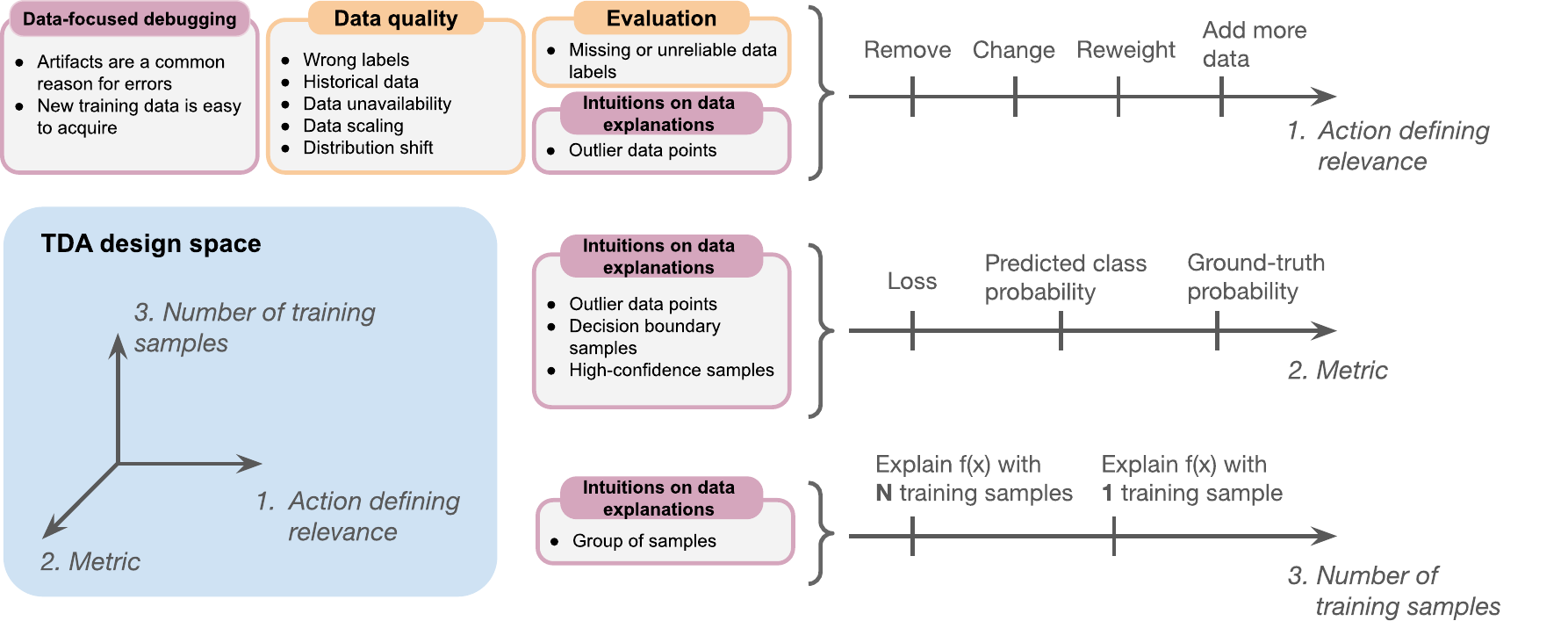}
    \caption{Three-dimensional design space of TDA explanations representing the type of data-centric information useful in model development. Each axis is derived from the themes and codes found in the interviews.}
    \label{fig:tda_design_space}
    \Description{The figure consists of two parts: One part shows three axes of a space titled TDA design space: 1. Action defining relevance, 2. Metric, 3. Number of training samples. The other part shows which themes and codes from the interview study informed each axis in isolation. The axes are stacked vertically, and the first axis is informed by the themes Data-focused debugging, Data quality, evaluation and Intuitions on data explanations. The second axis is informed by TDA intuitions and the third axis as well. }
\end{figure}

Drawing from our interview findings, we explore the design space of TDA explanations in model debugging. 
Interview participants provided rich information about what they usually look for in the training data when debugging a model, and for which purpose. Even though the objective and desired approach are similar, i.e., improving the model performance given a particular bug and data-centred debugging, respectively, developers tend to seek different information first. 
Specifically, we found that different data artifacts are investigated using different \textbf{actions}, with users interested in various \textbf{metrics} and considering different \textbf{numbers of training samples} depending on the context. From these observations, we define a three-dimensional design space (see Figure~\ref{fig:tda_design_space}). 
In the following, we detail how each axis was driven from the themes and codes found in the interview study (Figure~\ref{fig:interview_tdacodes}).

\subsubsection{Axis 1: Action defining relevance}

The first axis corresponds to the \emph{actions} participants propose for different types of data artefacts: Removing data, changing the label, reweighting data and collecting more data (Figure~\ref{fig:tda_design_space}). 
This axis is inspired by the participants' statements about the data-centric nature of development work (Theme: Data-focused debugging). Dataset curation and manipulation is a common activity in development work and has been described as the \textit{``[largest lever]''} (I1) for model performance. Different actions were mentioned or are implied in the interview study, which are linked to different hypotheses about the cause: 

Participants look at data to \textbf{\emph{remove}} when they believe that the model may have learnt the behaviour of interest from malicious data, for instance, mislabelled or outdated, historical data (Theme: Data-focused debugging $\rightarrow$ Code: Artifacts are a common reason for errors, Theme: Data quality $\rightarrow$ Codes: Wrong labels, Historical data). 
Additionally, we find that participants look for data that they wish to \textbf{\emph{change}}, for example filling in or imputing missing data, or correcting wrong data scales and labels (Theme: Data quality $\rightarrow$ Code: Data unavailability, Data scaling, Theme: Evaluation $\rightarrow$ Code: Missing or unreliable data labels). Especially when data is hard to obtain (e.g. in the medical domain), removing data is not an option and correcting data is preferred (P6). Another action is the \textbf{\emph{reweighting}} of training samples and \textbf{\emph{collecting more data}} related to the error which participants mentioned to be rather easy in some domains like autonomous driving (P1). These actions emphasise specific samples in the training process to encourage the model to learn them (Theme: Data-focused debugging $\rightarrow$ Code: New training data is easy to acquire). This is particularly relevant to learning outliers and handling distribution shifts (Theme: Data quality $\rightarrow$ Code: Distribution shift, Theme: Intuitions on data explanations $\rightarrow$ Code: Outlier data points). 

Removal, change and addition of data have different effects on a model and therefore represent different quantifications of attribution that answer different questions. For instance, the removal action implies a question like ``How would the model behaviour change if one were to exclude a training sample?'' which corresponds to traditional TDA (Equation~\ref{eq: loo} in \autoref{sec:tda}). The addition of training data, however, implies the question of ``How would the model behaviour change if one were to expand the dataset?'' which is a largely different question. By studying preferences in this axis, we aim to identify what data attribution needs to represent to be actionable and reflect the debugging process. 

\subsubsection{Axis 2: Metric}

The second axis corresponds to the sample-wise metrics participants used when referring to model behaviour: the loss, the probability of the true class label, and the probability of the predicted class label (y-axis of Figure~\ref{fig:tda_design_space}). This axis is inspired by the quantities participants talked about when explaining their intuitions about TDA (Theme: Intuitions on data explanations).

The interviews show that not all participants think of model behaviour in terms of sample loss, but also in classification probability. While high \textbf{\emph{loss}} can be an indicator for wrong or outlier samples (\textit{``[Samples with high training loss are] often [...] mislabeled examples.''} (I1), Theme: Intuitions on data explanations $\rightarrow$ Code: Outlier data points), participants also talked about relevant samples as those where the model has high or low classification \textbf{\emph{probability}}. Such samples provide context to the global model behaviour to the model developer, specifically indications about the decision boundary (\textit{``I would be interested in two kind of samples, the ones that [the model is] super confident with. But also the samples that are, I guess 50-50, it could be either one class [to see] why [the model] still chose class A.''} - I4, Theme: Intuitions on data explanations $\rightarrow$ Codes: Decision boundary samples, High-confidence samples) and can therefore be helpful for their work.  

The metric can be seen as a measure by which participants primarily perceive per-example model performance in the debugging process: The loss is indicative of the performance of the task, but also prediction probability interpreted as model confidence in a decision matters for understanding model behaviour. Both offer different types of information about the model.
By understanding metric preferences, we aim to identify the metric that helps model developers most in debugging and testing whether the loss, as in traditional TDA, is the most actionable measure. 

\subsubsection{Axis 3: Number of training samples}

The third axis corresponds to the number of training samples participants need to make sense of a model error through the training data. We identify two main options: A single sample and a group of samples (z-axis of Figure~\ref{fig:tda_design_space}). This axis is inspired by the explicit answer of I5 (\textit{``I think [the most relevant data] would be like a collection of [samples].''}) and the implicit indication of talking about multiple samples in the participants' answers, recognisable through the plural form use of \textit{``data''}, \textit{``samples''} and \textit{``artefacts''} (Theme: Intuitions on data explanations $\rightarrow$ Code: Group of samples).

We found that model developers usually do not inspect just one training data sample but think more globally in groups of relevant data. This aspect is addressed in TDA research by works studying group attribution~\cite{koh2019accuracy, basu2020second, park2023trak, hu2024most}. 
By understanding preferences in this axis, we aim to identify whether individual attribution, like in traditional TDA, or rather training data groups, needs to be in focus of TDA research.

\subsection{Scenario-based Interactive User Study}
\label{method:survey}

Building on the findings of the interview study, we designed an interactive online study that explores the design space of TDA explanations (see Figure~\ref{fig:tda_design_space}) 
using scenario-based design~\cite{carroll1995sbd}. We piloted the study interface to ensure clear phrasing and compatibility across devices. Data was collected from June -- August 2024. 

\subsubsection{Participants} 
\begin{figure}
    \centering
    \includegraphics[width=\linewidth]{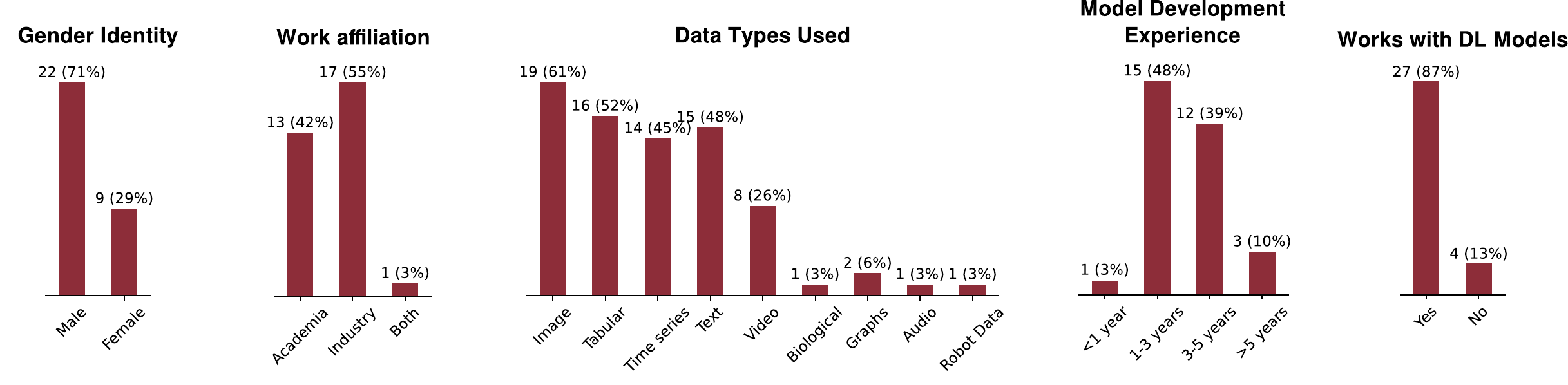}
    \caption{Histograms of answer distributions for the preliminary demographic questions of the scenario-based interactive study (N=31). DL = Deep learning.}
    \label{fig:survey_participants}
    \Description{This figure shows five histograms about the answer distributions for the preliminary demographic questions of the scenario-based, interactive study. From the top right corner to the bottom left corner, the histograms show the gender distribution (22 male vs. 9 female), whether people work in industry or academia (13 academia, 17 industry, 1 in both), years of experience distribution (1 <1 year, 15 1-3 years, 12 3-5 years, 3 >5years), the distribution of data types that participants work with (19 Image, 16 tabular, 14 time series, 15 text, 8 video, 1 biological, 2 graphs, 1 audio and 1 robot data), the distribution of experiences with deep learning models (27 yes vs. 4 no).}
\end{figure}

We recruited participants with prior experience in developing machine learning models to closely align with potential real users. Contrary to the interview study, we did not constrain our participants' work to be in a high-risk application domain for improved reach.
To recruit expert participants, we used snowball sampling alongside social media ads on X and LinkedIn, reaching out to personal contacts and asking them to share the study link. We aimed for around 30 replies to provide sufficient diversity in the answers for the qualitative analysis while also being indicative of measurable effects for a quantitative analysis.\footnote{We conducted a $\chi^2$ test of statistical significance. A power analysis at significance level $p<0.05$, power $\beta=0.7$ and large effect size $w=0.5$ yields a sample size of 31.}
Of 34 responses, we excluded three for low-quality answers (short and non-sensical), leaving 31 participants (nine female). 18 participants work in industry while 14 work in academia, with most having 1-5 years of machine learning experience. Four do not work with deep models. Participants come from diverse fields, including biology, health, logistics, and more. While the data types used by participants are diverse too, the majority of participants work with image or tabular data, as used in our study (see Figure~\ref{fig:survey_participants}). The participants were compensated with 25€ (=\$27.05) via Tremendous\footnote{\url{https://www.tremendous.com/}}.  

\subsubsection{Scenario-based design} 
\label{sec:scenario_desc}
\begin{figure}
    \centering
    \includegraphics[width=0.95\linewidth]{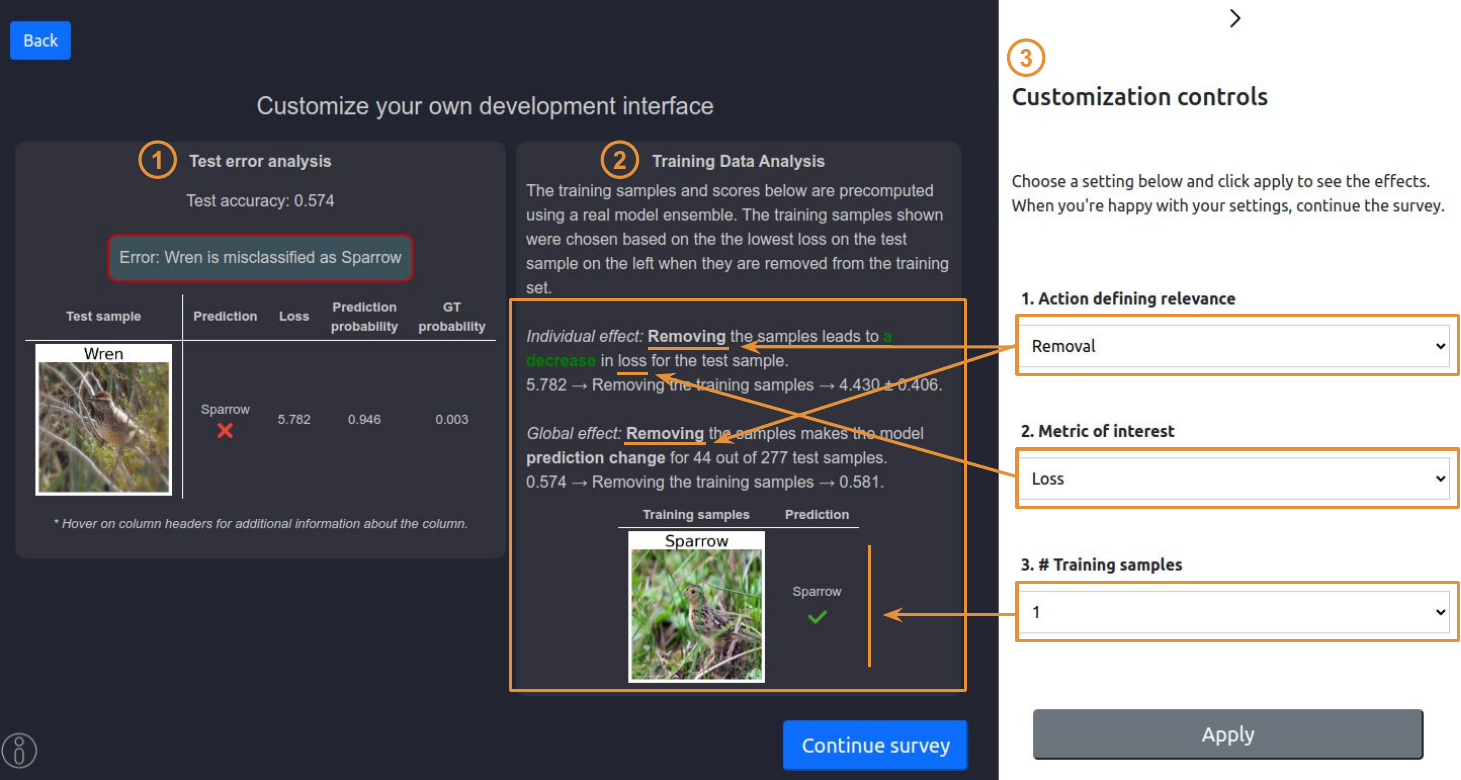}
    \caption{Screenshot of the interface used in the bird classification scenario. The dark part side shows the main interface consisting of the left panel (1) which shows information about the test error and is static, and the right panel (2) which shows information about the training data analysis and is dynamic. The information in (2) changes based on the chosen setting using the customisation controls (3) (expandable menu with light background).}
    \label{fig:screenshot_scenario}
    \Description{The figure shows a screenshot of the bird classification scenario interface of the survey study. The page is divided into two parts: On the left is the interface with the header ``Customize your own development interface'', and on the right are the customization controls as an expanded menu. The controls have the instructions: ``Choose a setting below and click Apply to see the effects.'' Below are three drop-down lists, one for Action defining relevance, one for Metric of interest and the last one for the number of training samples. The settings are set to ``Removal'', ``Loss'' and ``1'' respectively. Below the dropdown lists is a grey button with the word Apply on it. The interface itself is divided into two panels as well: on the left, indicated with a ``1'' is a panel with the title ``Test error analysis''. Below, the test accuracy is given as well as a description of the Error (``Wren is misclassified as Sparrow.''). Underneath the error description is the image of the wren that is misclassified with information about the prediction, loss, predictive probability and ground truth probability. The second panel, indicated with a ``2'' is titled ``Training data analysis''. It shows both the individual effect (how does the model behaviour on the erroneous test sample change?) of the setting chosen in the customisation menu as well as the global effect (how does the overall model behaviour change?). Below the effect is a table with the most relevant training samples according to the chosen setting, with their ground truth and predicted labels. The dynamic information is highlighted by an orange box with arrows pointing from the customisation menu to the dynamic panel.}
\end{figure}

The main part of the study is constructed with scenario-based design~\cite{carroll1995sbd} to base the analysis on a tangible usage scenario. We created two scenarios of model debugging in an interactive mock-up of a model development suite (see Figure~\ref{fig:screenshot_scenario}). The choice of data types (image and tabular data) was based on the dominant modalities of interview participants. The study put participants in two imaginary scenarios where they are model developers in a company that builds (1) a bird classification app, and (2) a credit scoring app. The company recently acquired a data-centric tool to help model developers debug their models by understanding errors and identifying training data relevant to those errors. The tool is customisable to adapt to the developer's preferences. 
The customisation choices, based on the TDA design space (see Figure~\ref{fig:tda_design_space}) were: 
\begin{enumerate}
    \item \textbf{Action defining relevance}: (a) Removal, (b) Label change (to the next most likely class), (c) Upweighting (implemented with a factor of 10). We omit new data collection because it is difficult to predict what kind of data participants would like to add. Instead, we take upweighting as an approximation of adding new data. 
    \item \textbf{Metric}: (a) Loss, (b) Ground truth class probability, (c) Predicted class probability.
    \item \textbf{Number of training samples}: Participants can choose between the 1 -- 10 training samples to inspect. We limit the number to 10 to explore preferences for more training samples while keeping computation efforts low.
\end{enumerate}

In both scenarios, participants were presented with a specific misclassification as the model error (left side of Figure~\ref{fig:screenshot_scenario}). This error was randomised across participants to prevent bias from a specific sample. Participants were asked to choose which type of training data-related information (right side of Figure~\ref{fig:screenshot_scenario}) is most useful to understand the reasons for the error. 
This way, we aimed to elicit participants' intuitions about the most actionable information for debugging the model error. Participants were encouraged to explore different settings and submit their desired customisation to complete the task. The data for these scenarios was generated using real machine learning models and openly available datasets~\cite{WahCUB_200_2011, hofmann1994statlog}. We detail the data generation process in 
 Appendix~\ref{appendix:datagen}. 

\subsubsection{Study structure} 
\begin{figure}
    \centering
    \includegraphics[width=0.95\linewidth]{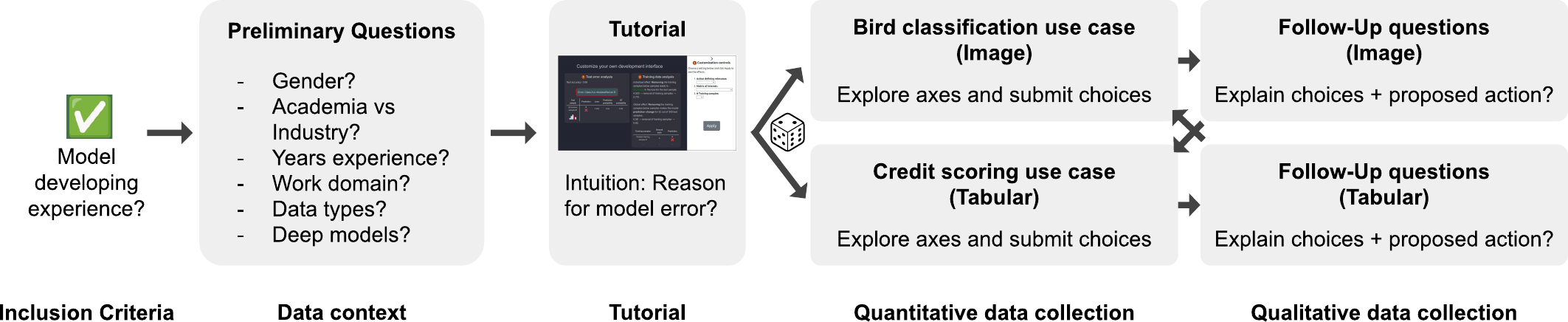}
    \caption{Structure of the scenario-based interactive user study. It is randomised which use case (i.e., image or tabular) is first presented to the participant after the tutorial to avoid potential biases in the answers arising from the order in which use cases are presented.}
    \label{fig:survey_overview}
    \Description{The figure shows a flowchart of the scenario-based interactive user study. The study starts with a question about prior model developing experience to assess whether the inclusion criteria are met. Then, the next state of preliminary questions follow to record context to the data. This includes demographic questions and questions about prior model development experiences. The third state is the tutorial for the interactive application, after which the quantitative data collection is started with either the bird classification (image) use case or the credit scoring (tabular) use case. In this stage, participants can explore the interface and are asked to submit their final choices. The next stage is the qualitative data collection, where participants are asked to explain their choices and proposed actions in free text. Then, an arrow indicates that the other use case will be presented before the end of the survey.}
\end{figure}

The structure of the scenario-based interactive study is shown in Figure~\ref{fig:survey_overview}. Participants were first briefed on the study's objective and data processing. After providing consent and confirming they meet inclusion criteria, they were asked preliminary demographic questions for context to the later analysis.
In the remaining study, we aimed to extract the participant's preferences through a usage scenario (\autoref{sec:scenario_desc}). The participants were presented with the scenario description that details their objective of understanding possible reasons for a presented error. They were prompted with ``Imagine you are a model developer in a company [...]'' to immerse them in the use case. 
Then, we asked the participants follow-up questions to understand the reasoning behind their choices through free-text questions (e.g.``Why did you choose to look at the \texttt{\{chosen\_metric\}} metric?''). 
These free-text questions were about understanding the participants' intuitions of why the model may have made an error (``In your opinion, why did the model fail at the above classification?''), and how they would fix it (``What would you do to fix this error?''). 
After completing one use case, the participants were presented with the other use case, where the order is randomised across users to prevent an order bias.

\subsubsection{Data analysis} 
We analyse the scenario-based interactive study both quantitatively and qualitatively. 
The quantitative analysis aims to examine the diversity of participant preferences in TDA explanations. Specifically, we investigated whether the predominant form of explanation in TDA, i.e. the counterfactual effect of removing a sample (\autoref{sec:tda}), aligns with the explanatory needs of model developers. If there is alignment, TDA research focuses on the right problem, and methods produced in this field are valuable to model developers. However, if this is not the case, it is a strong signal that the research should be diversified into multiple possible scenarios serving different user needs.

To this end, we computed the distribution of the participants' choices (action, metric and number of samples). We tested whether the collected data indicates a statistically significant preference for a specific setting. We performed a $\chi^2$ test per dimension of the design space (action defining relevance, metric, number of training samples) as the data is categorical~\cite{pearson1900x}. We considered two categories for the number of training samples, the first one being a single sample and the second one being groups larger than one as defined in the design space (Figure~\ref{fig:tda_design_space}). 
Specifically, our hypotheses are:  
\begin{align*}
    H_0 &: \text{Model developers have \textbf{diverse} needs for TDA explanations.}\\
    H_1 &: \text{Model developers need a \textbf{specific} definition of TDA.}
\end{align*}
In other words, we test the goodness of fit of our observations against a uniform distribution across categories of an axis. Since we conduct multiple tests on the same data, we apply a Bonferroni correction~\cite{weisstein2004bonferroni} and define statistical significance at $p<0.05$. 
Additionally, we compute each axis category's entropy $\mathbf{H(X)}$ and normalised entropy $\mathbf{H(X)}_{\mathrm{norm}}$ (normalised by $\log_2 (k)$ with $k$ as the number of categories in an axis, e.g. three for the metric axis) to get an indicator of the diversity of preferences~\cite{alsakran2014using}.
\begin{equation}
    \label{eq: entropy}
    \mathbf{H(X)} = \sum_{x\in\mathbf{X}} p(x)\log p(x)
\end{equation}
If the entropy is high, the participant's preferences are highly diverse, meaning there is no agreement on a certain preferred notion of the relevance of training data. 

\begin{figure}
    \centering
    \includegraphics[width=0.9\linewidth]{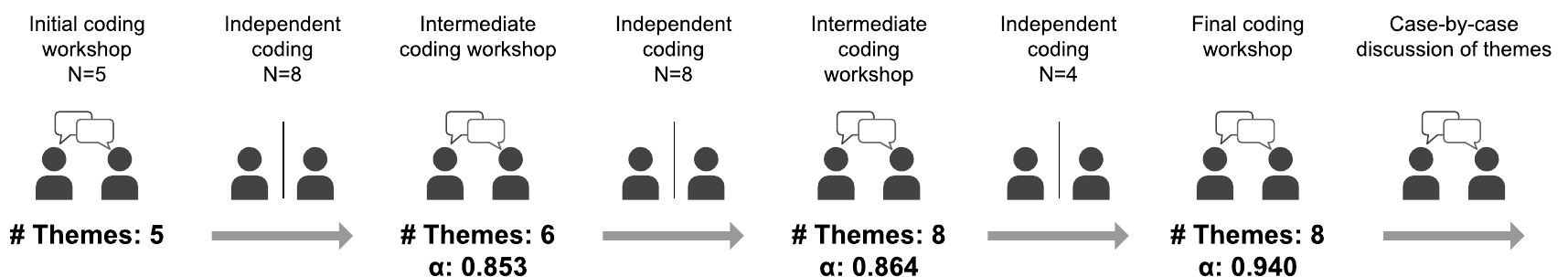}
    \caption{Iterative thematic analysis process showing the evolvement of themes and interrater agreement in Krippendorffs $\alpha$~\cite{krippendorff2011computing}.}
    \label{fig:coding_process}
    \Description{TODO}
\end{figure}

The qualitative analysis aims to understand the reasons behind participant choices and get deeper insight into user preferences. The free-text responses in the follow-up questions were analysed using inductive thematic coding~\cite{braun2012thematic} by two coders. Per question, we analysed the common themes among participants to extract the reasons behind user preferences and needs. The analysis consisted of several coding workshops and is depicted in Figure~\ref{fig:coding_process}. We measure interrater agreement in Krippendorff's $\alpha$~\cite{krippendorff2011computing}. The final agreement is $\alpha=0.94$.

\subsection{Study findings}
\label{results:survey}

We conducted the interactive study to study model developers' TDA needs through a scenario. The data analysis indicates a clear connection between developers' intuition of what causes a model error and their TDA needs. We elaborate further in the following. 

\subsubsection{Quantitative results}
The quantitative analysis builds on the customisation choices of the debugging interface (see Quantitative data collection in Figure~\ref{fig:survey_overview}). Table~\ref{tab:survey_preferences} shows the frequency of selections and the associated p-values. 
We test whether a specific action, metric, and number of training samples are consistently chosen and particularly useful to model developers. We are especially interested in a clear preference for removal, loss, and one training sample as these correspond to the predominant notion of TDA introduced in \autoref{sec:tda}. The analysis yields the following results: For the \textbf{action} axis, there is no statistically significant preference for any choice, and the distribution is rather uniform across choices. 
While the results may show a higher frequency of participants choosing loss as a \textbf{metric}, we observe no statistically significant preference. Hence, we fail to reject $H_0$ for the action and metric axes and conclude that there is no dominant preference for a particular TDA setting. Instead, developers have individual preferences. For the \textbf{number of samples} axis, however, we observe that all of our participants chose >1 samples across both scenarios. More specifically, we find a statistically significant preference for a group of 10 samples, indicating a clear preference for training data groups instead of individual samples.

The entropy analysis further supports this conclusion (see Table~\ref{tab:entropy_results}). As a measure of diversity, normalised entropy scores range from 0.588 to 0.997, showing a highly diverse (>0.5) distribution of choices. This is especially true for the \textbf{action} and \textbf{metric} axes with particularly high scores (>0.8). There is less diversity in the number of training samples but a clear preference for choosing groups of samples, confirming Ilyas et al.'s~\cite{ilyas2022datamodels} intuitions for studying group attribution. 

In conclusion, the quantitative analysis reveals that the preferred TDA explanations represent highly diverse information with a clear need for group attribution. 

\begin{table}[t]
    \centering
    \caption{Quantitative analysis of axes choices across participants in the frequency of choice (\%) for the image and tabular use case, as well as together. $p$ denotes the Bonferroni-corrected p-value computed from the $\chi^2$ test for statistical significance. GT=Ground truth.}
    \label{tab:survey_preferences}
    \small
    \begin{minipage}[t]{.45\linewidth}
        \centering
        \caption*{Action and Metric}
        \begin{tabular}{@{}p{0.24\textwidth}llllll@{}}
            \toprule
            & \multicolumn{2}{c}{Image} & \multicolumn{2}{c}{Tabular} & \multicolumn{2}{c}{Both} \\
            Choice & \% & $p$ & \% & $p$ & \% & $p$\\ \midrule
            \multicolumn{3}{l}{\textbf{Action}} \\ \midrule
            Removal  & 33.3\% & 1 & 42.0\% & 1 & 37.7\% & 1\\
            {Label Change}  & 30.0\% & 1 & 16.1\% & 1 & 23.0\% & 1\\
            Upweighting  & 36.7\% & 1 & 41.9\% & 1 & 39.3\% & 1\\ \midrule
            \multicolumn{3}{l}{\textbf{Metric}} \\ \midrule
            Loss  & 60.0\% & 1 & 35.5\% & 1 & 47.5\% & 1\\
            { Pred. cls. prob.} & 10.0\% & 1 & 41.9\% & 1 & 26.2\% & 1\\
            { GT class prob.} & 30.0\% & 1 & 22.6\% & 1 & 26.2\% & 1\\
            \bottomrule  
        \end{tabular}
    \end{minipage}%
    \hspace{0.03\linewidth} %
    \begin{minipage}[t]{.45\linewidth}
        \centering
        \caption*{Number of Samples}
        \begin{tabular}{@{}lllllll@{}}
            \toprule
            & \multicolumn{2}{c}{Image} & \multicolumn{2}{c}{Tabular} & \multicolumn{2}{c}{Both} \\
            Choice & \% & $p$ & \% & $p$ & \% & $p$\\ \midrule
            \multicolumn{3}{l}{\textbf{Num. of samples}} \\ \midrule
            1  & 0.0\% & 1 & 0.0\% & 1 & 0.0\% & 1\\
            >1 & 100.0\% & 0 & 100.0\% & 0 & 100.0\% & 0 \\
            \midrule
            2  & 0.00\% & 1 & 0.00\% & 1 & 0.00\% & 1\\
            3  & 3.33\% & 1 & 9.68\% & 1 & 6.56\% & 1\\
            4  & 6.67\% & 1 & 3.23\% & 1 & 4.92\% & 1\\
            5  & 13.33\% & 1 & 19.35\% & 1 & 16.39\% & 1\\
            6  & 3.33\% & 1 & 3.23\% & 1 & 3.28\% & 1\\
            7  & 0.00\% & 1 & 3.23\% & 1 & 1.64\% & 1\\
            8  & 3.33\% & 1 & 0.00\% & 1 & 1.64\% & 1\\
            9  & 3.33\% & 1 & 0.00\% & 1 & 1.64\% & 1\\
            10 & 66.67\% & <0.001 & 61.29\% & 0.002 & 63.93\% & <0.001\\            
            \bottomrule
        \end{tabular}
    \end{minipage}
\end{table}

\begin{table}[t]
    \centering
    \caption{Entropy ($\mathbf{H}$, defined in Eq.~\ref{eq: entropy}) and min-max normalised entropy ($\mathbf{H}_\mathrm{norm}$) per axis as a measure of diversity .}
    \label{tab:entropy_results}
    \small
    \begin{tabular}{@{}lllllll@{}}
        \toprule
        & \multicolumn{2}{c}{Image} & \multicolumn{2}{c}{Tabular} & \multicolumn{2}{c}{Both} \\
         Axis & $\mathbf{H}$ & $\mathbf{H}_{\mathrm{norm}}$ & $\mathbf{H}$ & $\mathbf{H}_{\mathrm{norm}}$ & $\mathbf{H}$ & $\mathbf{H}_{\mathrm{norm}}$ \\ \midrule
         Action & 1.095 & 0.997 & 1.023 & 0.931 & 1.073 & 0.976 \\
         Metric & 0.898 & 0.817 & 1.068 & 0.972 & 1.056 & 0.961 \\ 
         Num. samples & 1.173 & 0.603 & 1.176 & 0.657 & 1.224 & 0.588 \\ 
         \bottomrule
    \end{tabular}
\end{table}

\subsubsection{Qualitative results} 

\definecolor{surveyorange}{RGB}{249,203,156}
\definecolor{surveyblue}{RGB}{201,218,248}
\definecolor{surveygreen}{RGB}{182,215,168}
\definecolor{surveyred}{RGB}{230,184,175}
\definecolor{surveyyellow}{RGB}{255,229,153}

\begin{figure}[tp]
    \centering
    \includegraphics[width=0.95\linewidth]{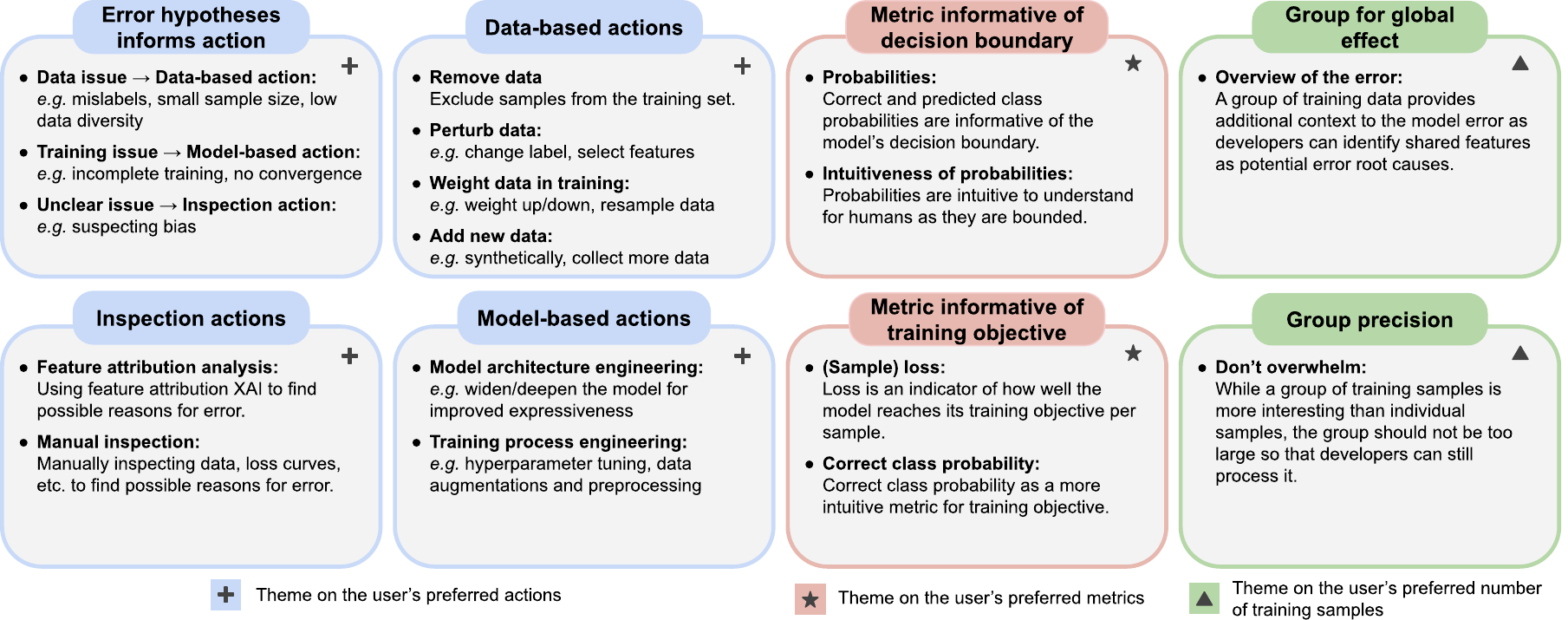}
    \caption{Themes with respective codes from the qualitative analysis of the survey.}
    \label{fig:survey_themes}
    \Description{This figure depicts the eight themes found from the qualitative analysis of the survey answers. There are four themes related to the participant's preferred actions (Error hypothesis informs action, Data-based actions, Inspection actions and Model-based actions), two themes related to the preferred metric (Metric informative of decision boundary, Metric informative of training objective) and two themes related to the preferred number of training samples (Group for global effect, Group precision).}
\end{figure}

The inductive thematic coding analysis of the free-text answers resulted in eight themes that address the reasoning behind participants' choices for \ulcolor[surveyblue]{action}, \ulcolor[surveyred]{metric}, and \ulcolor[surveygreen]{number of samples} preferences (see Figure~\ref{fig:survey_themes}). We elaborate on the themes below. 

\paragraph{\textbf{\ulcolor[surveyblue]{Theme: Error hypotheses informs action}}}
The thematic analysis complements our quantitative findings: Developer preferences in model debugging are highly individual. The analysis provides a reason in line with the sensemaking framework introduced in Cabrera et al.~\cite{cabrera2023what}: Individual preferences for proposed actions are often linked to the developer's understanding and intuitions. Needs are highly individual because developers may have different mental models based on their expertise and past experiences. In other words, their mental models of the model's learned behaviour affect their hypothesis about an error and the chosen action. The developer's hypothesis determines their explanatory needs: An explanation needs to enable the developer to reject or confirm their hypothesis of why a model error occurs so that they can make informed decisions and actions to resolve the error. 
For example, participants who hypothesise the cause of an error to be data-related also propose data-related actions: \textit{``[Images] [...] in the training set are not diverse enough''}, \textit{``[One class] was not representative enough''}. Corresponding to this hypothesis, participants often suggested adding new data or applying data augmentations to diversify the training set. If the hypothesis is linked to the model setup, the proposed action corresponds to the model.

\paragraph{\textbf{\ulcolor[surveyblue]{Themes about type of actions}}}
In terms of types of actions, we identified three main themes: Data-based actions, model-based actions, and inspection actions. 
In particular, the data-based actions are relevant to studying TDA needs (Theme: Data-based actions). In line with the interview study findings, we found that there are various data-based actions that participants propose: From removing malicious data to balancing the dataset through reweighting or adding new data. Additionally, we found that participants propose synthetically creating new data to add diverse data or fill up undersampled parts of the dataset (\textit{``address class imbalance by synthesizing data''}). This shows that the actions participants propose are various and our collection is not likely to be exhaustive. Hence, useful and actionable TDA needs to facilitate a variety of actions that help developers verify their hypotheses about the model error. 
Model-based actions address changes in model architecture or hyperparameter tuning and are connected to participant hypotheses about incomplete training (\textit{``Apply regularization methods like dropout or batch normalization to improve generalization and reduce overfitting.''}) (Theme: Model-based actions). 
Inspection actions are linked to first understanding a suspected bias of the model to get a clearer hypothesis about the error (Theme: Inspection actions). We found feature attribution explanation approaches (\textit{``Check which features the model considers important using techniques like SHAP [...].''}) and manual inspection-based approaches (``Typically, I would focus on the specific test sample and try to understand why the model is making a mistake by analyzing the feature values and their relationships and checking for outliers or anomalies.''). TDA explanations were not mentioned, but could ideally support these inspection activities. 

\paragraph{\ulcolor[surveyred]{\textbf{Themes about metrics}}}
We identified a link between the preferred metric and the task at hand. In the case of binary classification, participants showed a preference for probability metrics, as it is easier to understand than the loss and allows an inspection of the decision boundary (\textit{``The loss is not very intuitive''}, Theme: Metric informative of decision boundary). In multi-class classification, more participants preferred the loss as it considers all possible classes (\textit{``[The loss] contains information on all classes''}, Theme: Metric informative of training objective). However, as the quantitative analysis shows, these differences in preference are not statistically significant. Therefore, the metric preferences and actionability are likely individual to the user and their habits.

\paragraph{\ulcolor[surveygreen]{\textbf{Themes about number of training samples}}}
The quantitative study showed a clear preference for attribution to groups of training samples as opposed to individual training samples. The qualitative analysis found that this is due to the developer's need for having a good overview of the model error. Inspecting a group of training samples provides a better overview than a single sample, and allows the user to understand whether their hypothesis about the error is valid or not (Theme: Group for global effect). A second theme highlights another angle: While variety in the group is desired for a good overview of the error (e.g. group consisting of samples with diverse features), the group should be concise enough to \textit{``keep the analysis manageable''} and group size should be small enough to ensure reliable attribution scores, as changing the dataset could change the model's reasoning (Theme: Group precision).

In conclusion, the analysis of the scenario-based interactive user study deepens our understanding of model developers' needs for data-related explanations. We identify why there are diverse information needs across different developers: Since each developer may have a different mental model of their ML system, the information that is actionable to them is diverse as well and results in highly individual needs. 

\section{Discussion}
\label{sec:discussion}

This work presents a needfinding study for training data attribution (TDA) explanations in model development to inform user-focused TDA research in a top-down approach. We summarise our findings to provide answers to the research questions (\autoref{sec:rq_answers}), identify research gaps that need to be addressed for more user-focused TDA (\autoref{sec:futurework}). We discuss the study framework in a general sense as an approach to bridge method and needfinding research especially in early development phases of a technology (\autoref{sec:needframework}), followed by limitations (\autoref{sec:limitations}). 

\subsection{Answers to the research questions}
\label{sec:rq_answers}
This study sets out to understand user needs for TDA explanations in model development. We pose three questions to understand use cases, explicit and latent needs, and understudied areas of research in user-focused TDA. We provide our answers based on our study below.

\subsubsection{RQ1: What data-related explanation needs, both explicit and latent, do model developers have?}

To address this question, we conducted a two-part user study comprising an interview study (\autoref{method:interview}) and a scenario-based interactive study (\autoref{method:survey}). We aimed to explore how data-related explanations are used in model development and to gain a deeper understanding of both explicit and latent needs.

Explanations are needed when the model behaves unexpectedly. Since ML development is largely data-driven, developers often inspect unexpected model behaviour from a data-centric perspective (Interview theme: Data-focused debugging). We observed a strong connection between a developer’s mental model of the ML system and their corresponding explanation needs. In practice, developers tend to form hypotheses, often data-related, to make sense of unexpected model behaviour (e.g., a developer may hypothesise that a model behaves unexpectedly due to data artefacts). Explanations serve as a means to confirm or reject these hypotheses and to refine or generate new ones, and inform subsequent actions taken by the developer to improve the system (Interactive study theme: Error hypotheses inform action).

Since each developer’s mental model is shaped by their experience and expectations, their explanation needs are also diverse. The needs TDA explanations must meet reflect developers' data-centric information needs during model building. For example, developers seek insights to identify low-quality data when they suspect wrong labels to be the cause for a model error, or detect broader issues like distribution shifts when they believe this could have occured (Interview theme: Data quality), or recognise gaps where additional data is needed when they suspect their dataset to be insufficient for the model to learn the task (Interview theme: Data is most important). 

Hence, these findings reveal that model developers have explicit needs for explanations that help identify possible causes of model error, whether by examining the impact of removing certain training samples or by highlighting the potential value of adding new data. At a deeper level, developers also express latent needs: explanations must be actionable, reliable, and aligned with their mental models. Explanations are not only used to understand the model’s behaviour, but also to guide meaningful decisions during development.

\subsubsection{RQ2: To what extent do existing TDA approaches align with model developers' needs?}

To answer this question, we reflect on existing TDA approaches in light of the identified explicit and latent needs: TDA explanations need to be actionable, reliable, and aligned with the developer's mental models. From these needs, we identify different TDA tasks that developers need TDA explanations to fulfil, next to the overarching need for reliability (see Table~\ref{tab:attribution_types}). 

\begin{table}[ht]
    \centering
    \caption{Overview of the types of TDA identified in our study and how often they were mentioned in the survey study, defined by their specification in the TDA design space, where \emph{Attributing errors} was mainly mentioned in the additional comments. Cited works explicitly address this type of TDA. Elaboration on the cited works is provided in Appendix~\ref{appendix:extended_background}.}
    \label{tab:attribution_types}
    \small
    \begin{tabular}{@{}llllll@{}}
        \toprule
         \textbf{Attribution type} & \textbf{Action} & \textbf{Metric}& $\mathbf{|Z_{\mathrm{train}}|}$ & $\mathbf{|Z_{\mathrm{test}}|}$ &  \textbf{Times chosen} \\ \midrule
         Influence attribution (e.g.,~\cite{kohliang2017, pruthi_estimating_2020, guo-etal-2021-fastif, schioppa2022arnoldi, grosse2023studying, park2023trak, bae2024training}) & Removal & Sample loss & 1 & 1 & 0 \\ 
         Group influence attribution (e.g.,~\cite{park2023trak, basu2020second, koh2019accuracy, lin2024efficient}) & Removal & Sample loss & >1 & 1 & 11 \\ 
         Attributing errors & Any & Any & >1 & >1 & 4\\ 
         Lacking data attribution & Addition & Any & >1 & >1 & 26\\ 
         Decision boundary attribution & Label change & Any & >1 & >1 & 14\\ 
         \bottomrule
    \end{tabular}
\end{table}

\paragraph{Existing TDA approaches partly address explicit needs.}
The formal definition of TDA (\autoref{sec:tda}) characterises attribution as the change in model output resulting from training with versus without a specific training sample~\cite{hampel1974}. \cite{kohliang2017, grosse2023studying, park2023trak, bae2024training, choe2024your, kwon2023datainf, pruthi_estimating_2020} focus on efficiently and faithfully estimating this quantity (\textit{influence attribution} in Table~\ref{tab:attribution_types}) or its extension to groups of training samples~\cite{basu2020second, park2023trak, koh2019accuracy} (\textit{group influence attribution} in Table~\ref{tab:attribution_types}).
TDA has been applied to a variety of tasks, including the identification of mislabeled data, analysis of adversarial vulnerabilities, and detection of domain mismatches~\cite{kohliang2017}. Mislabel detection, in particular, is a widely used evaluation across several studies~\cite{pruthi_estimating_2020, park2023trak, schioppa2022arnoldi}, and TDA methods have demonstrated the ability to identify outliers and mislabeled samples in training data~\cite{kohliang2017, pruthi_estimating_2020}. These capabilities align with certain explicit developer needs, particularly those focused on identifying low-quality data to remove.
However, since TDA quantifies a sample’s relevance through the lens of its removal, it is especially well-suited for scenarios involving data elimination, such as detecting harmful or noisy samples. In contrast, broader data-centric explanation needs, such as identifying distribution shifts or detecting underrepresented regions in the data, are generally not addressed by current TDA methods, as they are not typically designed with these needs in mind.

\paragraph{Existing TDA work does not consider developers' latent need for actionable information.}
Our study shows that developers have the latent need for data-based explanations to enable them to confirm or revise their mental models of model behaviour, which are diverse. Fulfilling this latent need is tied to the developer's mental model and makes an explanation actionable. This human-centred aspect of actionability is rarely considered in existing TDA work. While existing TDA methods are useful for measuring the impact of removing and retraining on individual samples, because they are studied to quantify this objective, developers may require different forms of attribution to address a wider range of questions during model development (e.g., \textit{lacking data attribution} in Table~\ref{tab:attribution_types}). 

\paragraph{Existing TDA work partly implicitly addresses developers' need for reliability.}
Regarding the need for reliable information, existing TDA methods have been criticised for their low reliability in deep learning~\cite{nguyen2023bayesian, basu_influence_2021, bae_if_2022}. In addition, our study introduces a different aspect of reliability that only becomes apparent when considering the full context of the use case (as we assessed participants within the context of a specific task). When a dataset is modified, the overall model behaviour can be affected and must be considered. For TDA, this means that when modifying the dataset to address a particular error, developers require explanations that not only suggest an effective modification but also ensure the model's performance remains rather consistent. 

Our analysis underscores the importance of incorporating user perspectives to identify needs critical to the real-world application of this technology. Traditional TDA explanations only partially meet these needs, and further interdisciplinary efforts beyond method research are required to develop methods with improved actionability.

\subsubsection{RQ3: What kind of information produced by TDA will be useful for model developers?}
To answer this question, we examine the gap between the needs we found and the needs that existing TDA work meets (see Table~\ref{tab:attribution_types}). 

Our study reveals that developers need different types of information to understand model behaviour, depending on their mental models and hypotheses of why the model errs during development, in line with~\cite{cabrera2023what}. This necessitates TDA explanations that are flexible and adaptable to individual needs. These needs vary based on the application domain and the developer's mental models that are shaped by the developer's experiences and expectations. Experienced developers may form hypotheses about model errors based on their past experiences. In contrast, less experienced developers may form hypotheses based on what they learnt previously about machine learning models. TDA that aligns with the developer's mental models will be useful.

Developers prefer group attribution, as groups provide more information than individual samples and are expected to paint a more conclusive picture of the model behaviour. Hence, information related to group attribution will be useful for model developers. 

Additionally, developers need TDA explanations to be reliable, considering how data-centric actions might affect the model as a whole. Reliability entails two main aspects: First, the information needs to be reliable for it to be actionable and effective for the task (i.e., model debugging). Second, TDA explanations should also account for any broader changes in the model's behaviour that may affect the attribution to training samples.

\subsection{Overlooked research topics for user-focused training data attribution}
\label{sec:futurework}

This work adds to a collection of human-centred explainable AI (HCXAI) works calling to centre XAI research around users (e.g.,~\cite{brennen2020what, kaur2020interpreting, rutjes2019considerations, ehsan2024who, mei2023users, lin2024efficient}). We focus particularly on model developer users, as they have agency over the dataset. By focusing on TDA explanations specifically for model developers, we identify several research areas that are necessary to study to advance TDA research towards a focus on developers. 

\begin{itemize}
    \item \textbf{Mental models and TDA:} Our study supports previous work~\cite{rutjes2019considerations, kaur2020interpreting} that the most useful explanations depend on the user's mental model of why the model errs: We find that what information is \emph{actionable} to users depends in their hypothesis for reasons of model behaviour (Theme: Error hypothesis informs action). Therefore, future research should explore how TDA influences and is influenced by mental models, as well as how TDA can be adapted to reflect mental models (e.g., by defining relevance through the action of upweighting as opposed to removal), potentially leading to more aligned and actionable explanations that are well-defined.
    \item \textbf{User-centred group attribution:} The survey findings reveal a clear need for attributing model behaviour to groups of training samples, as it provides more informative insights (Theme: Group for global effect). While existing approaches~\cite{basu2020second, park2023trak, koh2019accuracy} address this, our findings highlight the need for creating groupings that \emph{make sense to users} (e.g., based on ground truth or predicted class), an aspect not yet widely explored in TDA research.
    \item \textbf{Holistic understanding of model errors:} Our thematic analysis of survey responses highlights the importance of understanding the model error itself before attempting to fix it (Theme: Inspection actions, Group precision). If the root cause is a spurious correlation, the error likely extends beyond the current test sample and may not fully represent the issue. We recommend future research focus on understanding error types rather than isolated instances (e.g., slice discovery~\cite{wang2024error}) and on attributing model behaviour on groups of test samples to training data (e.g.,~\cite{jain2024data}).
    \item \textbf{Reliability of TDA:} TDA quantifies the impact of training samples on model errors by predicting changes in model behaviour, serving as an explanation to the user. For the explanations to be informative, the attribution must be reliable. It is essential to understand how overall model behaviour will be affected, as mentioned by our participants. We recommend future research to extend existing work on the fragility, reliability, and stability of TDA~\cite{basu_influence_2021, nguyen2023bayesian, epifano_revisiting_2023, bae_if_2022} to address these needs.
    \item \textbf{TDA and feature attribution:} Combining feature attribution with TDA could create TDA explanations familiar to model developers who already use feature attribution methods like SHAP~\cite{lundberg2017unified}. While our participants indicated they frequently use feature attribution, there is a known risk of misinterpretation and overreliance~\cite{kaur2020interpreting, kim2022hive}. Providing feature attribution explanations with context, such as relevant training data identified through TDA, could help in checking whether similar features in training and test data might mislead the model. 
\end{itemize}

\subsection{A framework for needs-based research}
\label{sec:needframework}

In this work, we demonstrate a three-step process to identify potentially overlooked research directions and inform needs-based research for socially situated technology like machine learning model explanations: 
\begin{enumerate}
    \item \textbf{Understanding the user:} The first step aims to identify the potential user of a technology: Examining who may have an interest in the technology under consideration of their agency, as well as what purpose the technology serves. 
    In this work, we consider XAI technology. Hence, potential users interact with machine learning systems and have an interest in understanding them. Often in XAI, the users of ML technology in high-risk application areas are selected as the main audience due to legal requirements of transparency by the EU AI Act~\cite{aiact} for example, but interested users could extend beyond this group to any interested users. Since we specifically focus on TDA explanations, we were able to narrow the user group to model developers who have agency over the training data. 
    \item \textbf{Understanding their needs:} The second step aims to understand the needs of users through qualitative methods like semi-structured interviews with thematic coding. By engaging with users, we can understand the application context of the technology, including existing workflows that the technology needs to fit in and existing challenges that the technology could potentially help with. Especially in the case of novel technology that is not yet used, it is important to identify potential use cases to understand what needs exist. 
    In this work, we were able to analyse common use cases of TDA explanations in model development from interviews with machine learning developers (\autoref{method:interview}) and derive the design space of TDA explanations (\autoref{sec:design_space}).
    \item \textbf{Validate their needs:} The third step aims to validate the needs found in the previous step through a prototype or scenario-based study, so that users can experience the technology. This step enables analysis that is rooted in real interaction, and shall provide deeper insights and validation of the user needs. 
    In this work, we designed a mock-up interface for the scenario-based interactive user study and were able to deepen our understanding of what kind of information is needed from TDA explanations (\autoref{method:survey}). We were able to clearly identify preferences for group attribution, as well as currently understudied research directions, such as the need for explanations of model behaviour that point to missing data. 
\end{enumerate}

This process is intended to build a bridge between user research and methods research, as it allows for potential users to understand the capabilities of novel methods, and results in open questions and example use cases for method researchers. While this paper applies the framework to XAI, in particular TDA explanations, it is a general approach that could be applied to any technology intended for use in a sociotechnical context. 

\subsection{Limitations}
\label{sec:limitations}
This study has several limitations. First, the use of purposive and snowball sampling techniques in participant recruitment may have introduced selection bias. Consequently, the sample may not be fully representative of the broader population of professionals working with machine learning or model developers in general. 
Second, the relatively small sample size in our studies limits the robustness of the quantitative analyses, so the results should be interpreted with caution and lead us to focus more strongly on the qualitative analysis. 
Third, we simulated the use of TDA explanations in a model debugging interface, which are lab-like conditions for the survey study. Hence, the needfinding analysis is not rooted in the real workflows and is bound to our study conditions, including the interface design. 

Yet, we remark that this third limitation is somewhat systematic, as TDA explanations are not widely adopted in real workflows yet. At the time of the studies, our participants were not familiar with TDA explanations at all, and the first libraries for data attribution were only recently published~\cite{bareeva2024quandainterpretabilitytoolkittraining, deng2024texttt}. We therefore believe that future studies will overcome this limitation as more real-world users engage with the TDA pipeline in their work. 

\section{Conclusion}
\label{sec:conclusion}

This paper proposes a top-down approach to conducting explainable AI (XAI) research which is driven by user needs as opposed to techno-centric solutions. We illustrate this approach with the emerging subfield of training data attribution (TDA) and present a needfinding study with AI practitioners. The study consists of two stages: First, we conduct semi-structured interviews with machine learning developers in high-risk domain areas (N=6) to gain an understanding of common use cases and challenges. We derive a design space of TDA explanatory information that represents what relevant data for model debugging could look like. Second, we develop an interactive interface to explore developer preferences in the design space through a scenario-based interactive user study (N=31) targeted at model developers. The second study allows us to validate and expand the needfinding analysis. The mixed-methods analysis shows that the explanatory information needed from TDA for model debugging is highly dependent not only on the use case but also on the developer and their mental model and intuitions. Finally, we establish our study approach as a framework for needs-based research and highlight several research directions in user-focused TDA that are largely overlooked in the current landscape.

\begin{acks}
     We first and foremost thank all study participants who gave us their time and input. We also thank Albert Catalán, Nikita Kister, and Arnas Uselis for participating in the pilot study and helping us improve the study design. We thank Dustin Theobald for sharing his survey website codebase which we used as a basis. We thank Kristina Kapanova for helping us host the survey securely, and everyone who helped spread the survey study. The authors thank the International Max Planck Research School for Intelligent Systems (IMPRS-IS) for supporting Elisa Nguyen. This work was supported by the T\"ubingen AI Center.
\end{acks}

\bibliographystyle{ACM-Reference-Format}
\bibliography{references}

\include{appendix}

\end{document}

%% file: appendix.tex
\appendix

\section{Inclusion criteria: High-risk application areas}
\label{appendix:high-riskapps}
We refer to the definition of high-risk application areas according to Annex III of the European Union's AI Act~\cite{aiact}. For readability, we include an overview: 
\begin{itemize}
    \item AI applications in products that require a specific level of safety: 
    \begin{itemize}
        \item Toys,
        \item Aviation,
        \item Cars,
        \item Medical devices,
        \item Lifts.
    \end{itemize}
    \item Biometric identification and categorisation of natural persons.
    \item Management and operation of critical infrastructure.
    \item Education and vocational training.
    \item Employment, worker management and access to self-employment.
    \item Access to and enjoyment of essential private services and public services and benefits.
    \item Law enforcement.
    \item Migration, asylum and border control management.
    \item Assistance in legal interpretation and application of the law.
\end{itemize}

\section{Data generation for the scenario-based interactive user study}
\label{appendix:datagen}

The second probe of our study presents an interactive mock-up of a model development suite where participants can customise what kind of data-centric information is shown. This data is pre-computed using real models and openly available datasets for an image classification and a tabular data classification use case. In the following, we describe the data generation process. 

\subsection{Use case: Bird classification app}

The setting of the image classification use case is about a company that builds a bird classification app. This bird classification app setting has been used in previous HCXAI studies, e.g. \citet{kim2023help}. To generate the data for this use case, we finetune the pretrained ResNet18~\cite{he2016deep} checkpoint in PyTorch~\cite{pytorch} with the CUB dataset~\cite{WahCUB_200_2011}. 

\paragraph{Data preprocessing.} The CUB dataset~\cite{WahCUB_200_2011} is a multi-class image classification with 200 classes, and 5994 images in training and 5794 images in the test set. While the dataset includes fine-grained annotations, we only utilise the target labels in the data generation process. We subsample CUB to include only 10 classes for our application to ensure that participants can learn and identify the different bird species without prior knowledge. We refer to the selected classes as their species family (e.g. Grasshopper sparrow $\rightarrow$ Sparrow) for brevity. The training set includes 300 images, and the test set 277 images. 

\paragraph{Model training.} We finetune a ResNet18 model pretrained on ImageNet (available on the torchvision hub) with the CUB training subset for 10 epochs using the AdamW optimizer, a learning rate of 0.001 with weight decay factor 0.005 for the cross-entropy objective. Since the training process of deep models is stochastic which affects any retraining-based effects~\cite{nguyen2023bayesian}, we record the last three model checkpoints for a model ensemble. The final model achieves a test accuracy of 0.57 which is suitable for the model debugging use cases in this work. 

\subsection{Use case: Loan approvement recommendation app}

The setting of the scenario of the tabular data use case is the model development department of a bank that implements a machine learning model to give recommendations for loan approvals of their clients. To build this scenario, we train a logistic regression model on the German credit dataset~\cite{hofmann1994statlog} which is openly available on the UCI dataset repository and is licensed by CC BY 4.0. 

\paragraph{Data preprocessing.} The German credit dataset~\cite{hofmann1994statlog} is a multivariate dataset with 20 features and 1000 entries. It is used for binary classification and includes labels of loan approval and declines. We split it into a train and test set using an 80-20 train-test split. As we wish to display the whole data in the interface mock-up, we subselect the following features for readability: 
\begin{enumerate}
    \item The \textbf{status of existing checking account} feature describes how much money is in the applicant's checking account and is a categorical variable. Possible values are: (a) less than \$0, (b) less than \$200, (c) more than \$200, and (d) no checking account.
    \item The \textbf{duration} describes the loan's duration in months and is a numerical variable. 
    \item The \textbf{credit history} describes the applicant's history with loans from this and other banks as a categorical variable. Possible values are: (a) no credits taken, (b) all credits at this bank paid back duly, (c) existing credits paid back duly till now, (d) delay in paying off in the past, and (e) critical account.
    \item The \textbf{credit amount} describes the size of the loan in \$ and is a numerical variable.
    \item The \textbf{installment rate} in percentage of the applicant's disposable income describes how large the installment payments of the loan are. It is a numerical feature.	
    \item The \textbf{age} is a numerical feature and describes the age of the loan applicant.
\end{enumerate}

\paragraph{Model training.} We trained a logistic regression model with stochastic gradient descent using the scikit-learn library in Python~\cite{scikit-learn}. In detail, the model was trained in 50 epochs with L2 regularisation weighted by $\alpha = 10e^{-4}$ and the default adaptive learning rate of the library. Recording three checkpoints along the training trajectory at 30, 40 and 50 epochs results in the model ensemble we use for computing the data. The final model accuracy is $0.68$, leaving room for improvement and debugging. 

\subsection{Model error selection}

In the survey probe, the participant is asked to analyse a model error, i.e. a common misclassification. We select the errors to show by first understanding which type of misclassification is most common, i.e. which class A is often classified as another class B. To determine the test errors, we classify the test set using the model ensemble, group test samples according to true and predicted labels and select the errors represented by the largest groups. We select three error types for the multi-class problem in the bird classification use case. In the binary tabular classification use case, only two errors are possible. In these error groups, we randomly choose three erroneous samples each, resulting in nine and six erroneous test samples for the bird classification and loan approval recommendation use cases, respectively. The test samples are randomised across participants by drawing from a uniform distribution to prevent bias in the analysis stemming from a specific error or test sample. 

\subsection{Relevant training sample selection}

In the mock-up interface used in the survey probe, we display the effect of different training data-based actions on the model error at hand and the global model behaviour. We precomputed these values using the models presented in the above. We elaborate on the procedure of precomputation below. 

\paragraph{Determining the training samples.} The interface shows training samples most relevant to the test sample group based on different actions, i.e. removal, label change and upweighting, measured in different metrics and for different group sizes. To determine which training samples to show, we would have to precompute all possible combinations of action, metric and group size. As this is computationally prohibitive and rigorous correctness is not required for our research questions, we select the relevant training sample groups to the erroneous test samples based on individual relevance. In other words, we first iteratively remove/change the label/upweigh each training sample and retrain the models (details below). Then, we record the differences in loss/true class probability/predicted class probability for each erroneous test sample example and compute mean, variance and p-values as in \cite{nguyen2023bayesian}. After filtering the training samples that exhibit a statistically significant effect ($p<0.05$), we sort the training sample means from most positive to most negative effect which differs across metrics (e.g. an increase in loss is negative, while an increase in true class probability is positive). The top 10 training samples in each setting are the relevant training samples we use in the interface. 

\paragraph{Action implementation.} Implementing and computing the actions requires elaboration:
\begin{itemize}
    \item \textbf{Removal}:The implementation of removal is rather straightforward -- we exclude the training sample from the dataset and retrain using the same training pipeline.
    \item \textbf{Label change:} Changing the label of a training sample makes sense when another label may fit the sample better. Hence, we flip the label in the case of binary classification. In the case of more than two classes, we consider two cases: (1) The training sample was misclassified. Assuming that the model learns all classes well, a misclassification could occur because the predicted label describes the data better. Therefore, in these cases, the new label is the predicted label. (2) The training sample was correctly classified. We would only want to change the label if it was wrongly labelled. Assuming that the model generally learned the classes well, we changed the label to the predicted label with the second highest probability. 
    \item \textbf{Upweighting:} We implement upweighting with a weight factor of 10.
\end{itemize}

\paragraph{Computing final effects.} In the interface, we provide simulated effects of certain data-centric actions on the model errors (e.g. removing certain training datapoints leads to an increase in loss). Individual effects do not necessarily add up to the group effect~\cite{basu2020second, koh2019accuracy}. Hence, we compute the final effects using the actual removal/label change/upweighting of training data groups. This computation is feasible since the groups and the number of times we have to retrain the models are limited. We compute and show the standard deviation of removing/changing the label/upweighting across the model checkpoints recorded.

\section{Extended background: Training Data Attribution Methods}
\label{appendix:extended_background}

We provide a high-level overview of TDA methods cited in Table~\ref{tab:attribution_types} of the main paper and how they relate to the attribution types. Specifically, we elaborate on influence functions methods (\autoref{sec:if}), methods that trace the training process (\autoref{sec:tracin}), methods that unroll the training process (\autoref{sec:unrolling}) and group influence methods (\autoref{sec:group_if}). 

\subsection{Influence Function Methods}
\label{sec:if}

Influence functions (IF), first introduced in robust statistics~\cite{hampel1974}, are one of the main types of methods in TDA. Instead of costly retraining without a training sample $z_{\mathrm{train}} = (x_{\mathrm{train}}, y_{\mathrm{train}})$ to quantify $z_{\mathrm{train}}$'s influence on the model's behaviour $f_\theta(x_{\mathrm{test}})$ via Equation~\ref{eq: loo}, (IF) approximate this quantity by:    

\begin{equation}
    \label{eq:influence_function}
    \tau(z_{\mathrm{train}}, z_{\mathrm{test}}; f_\theta) \approx \tau_{\mathrm{IF}}(z_{\mathrm{train}}, z_{\mathrm{test}}; f_\theta) = -\nabla_\theta \mathcal{L}(f_\theta(x_{\mathrm{train}}, y_{\mathrm{train}}))^\top \mathbf{H}^{-1}\nabla_\theta \mathcal{L}(f_\theta(x_{\mathrm{test}}, y_{\mathrm{test}})) 
\end{equation}

where $\tau_{\mathrm{IF}}$ uses a Newton-like step to estimate the effect of a small perturbation on the model’s parameters, where the perturbation is the exclusion of $z_{\mathrm{train}}$. For a full derivation, we refer to \citet{kohliang2017} or \citet{bae_if_2022}. Intuitively, influence functions approximate the parameter landscape around the parameters $\theta$, and compute the update step taken in the direction of removing $z_{\mathrm{train}}$, to estimate its effect on the loss at $z_{\mathrm{test}}$. 

\paragraph{IF require approximations to be applicable to deep models.} 
Since equation~\ref{eq:influence_function} requires the inversion of the parameter Hessian ($\mathbf{H}^{-1}$), it also requires the objective function (i.e., the loss) to be twice-differentiable and the Hessian $\mathbf{H}$ to be positive definite, to ensure that $\mathbf{H}^{-1}$ exists. However, these conditions generally do not hold for deep models. Moreover, the Hessian $\mathbf{H}$ is a large and costly matrix to invert, so IF algorithms usually do not explicitly estimate $\mathbf{H}$ but implicitly through a matrix-vector product. They estimate the damped inverse Hessian Vector Product (iHVP) $(\mathbf{H}^{-1} + \lambda \mathbf{I})\nabla_\theta \mathcal{L}(f_\theta(x_{\mathrm{test}}, y_{\mathrm{test}}))$, where $\lambda$ is the damping parameter and $\mathbf{I}$ is the identity matrix. Damping improves the numerical stability of the iHVP estimation. In the following, we use the terms \textit{damped iHVP} and \textit{iHVP} interchangeably. 

\paragraph{IF methods generally differ in iHVP approximation.}
Different TDA methods propose different approximations of the iHVP: 
\citet{kohliang2017}, who first presented IFs for deep models, estimate the iHVP using an iterative estimator called \textit{Linear time Stochastic Second-Order Algorithm (LiSSA)}~\cite{agarwal2017second}. 
\citet{guo-etal-2021-fastif} also use LiSSA in their algorithm, but propose an additional speed up for the task of identifying the most influential training points for a test point by first retrieving candidate $z_{\mathrm{train}}$ with a k-nearest neighbour search in the model's embedding space. \citet{wang2025better} improve the iHVP and therefore influence estimation by using the Eigenvalue-corrected Kronecker-Factored Approximate Curvature (EKFAC)~\cite{grosse2023studying} as a preconditioning matrix on the iterative estimator. 
Alternatively, \citet{schioppa2022arnoldi} propose to speed up the approximation of the iHVP by projecting the computation into a lower-dimensional space spanned by the dominant eigenvectors of $\mathbf{H}$ with Arnoldi iterations~\cite{arnoldi1951principle}. Also \citet{park2023trak} and \citet{choe2024your} leverage a projection into lower dimensions to speed up the iHVP estimation, where the former uses random projections while the latter makes use of gradient structures. 
\citet{grosse2023studying} propose to estimate the $\mathbf{H}$ utilising Eigenvalue-corrected Kronecker Factored Approximate Curvature (EKFAC), allowing for a faster computation of the influence function since it is not an iterative method.

Influence function research is active and has been gaining traction in recent years. The main focus lies on finding an accurate estimation of the \textit{loss} difference on a \textit{single} test sample when a \textit{single} training sample is \textit{removed}. 

\subsection{Methods That Trace The Training Process}
\label{sec:tracin}

In 2020, \citet{pruthi_estimating_2020} presented a different way to compute training data attribution that traces a training sample's contribution to the model performance on a test sample throughout training. The objective definition of TDA is slightly different from the change after leave-one-out retraining (Eq.~\ref{eq: loo}): 
\begin{equation}
    \label{eq:tracin_ideal}
    \tau_{\mathrm{TracInIdeal}}(z_{\mathrm{train}}, {z_\mathrm{test}}) = \sum_{t;z_t = z_{\mathrm{train}}} l(\theta_t, z_{\mathrm{test}}) - l(\theta_{t+1}, z_{\mathrm{test}})
\end{equation}
TDA is hence defined as the sum of the loss difference on $z_{\mathrm{test}}$ at each training step with $z_{\mathrm{train}}$ across the full training. 

In practice, computing this idealised notion of TDA is also prohibitive across all training samples, so \citet{pruthi_estimating_2020} propose to approximate equation~\ref{eq:tracin_ideal} with a sum of gradient dot products across $k$ model checkpoints recorded during training:
\begin{equation}
    \label{eq:tracin}
    \tau_{\mathrm{TracInIdeal}}(z_{\mathrm{train}}, {z_\mathrm{test}}) \approx \tau_{\mathrm{TracInCP}}(z_{\mathrm{train}}, {z_\mathrm{test}}) = \sum_{i=1}^k \eta_i \nabla l(\theta_{t_i}, z_\mathrm{train})^\top \nabla l(\theta_{t_i}, z_\mathrm{test})
\end{equation}
where $\eta_i$ is the learning rate at checkpoint $i$. We note that in principle, each dot product corresponds to the formulation of influence functions (Eq.~\ref{eq:influence_function}) where the inverse Hessian is approximated by the identity matrix. 

TracIn~\cite{pruthi_estimating_2020} offers an easy-to-compute TDA method and objective. It is focused on the relationship between a \textit{single} training and test sample, measured on the \textit{loss} changes across training, whenever the training sample is seen during training. 

\subsection{Methods That Unroll The Training Process}
\label{sec:unrolling}

In a similar yet different spirit to tracing the training process as in TracIn~\cite{pruthi_estimating_2020}, another line of approaches to computing TDA are methods that are based on unrolling the training process~\cite{hara2019data, bae2024training, wang2025capturing, ilyas2025magic}, sometimes also referred to as dynamic TDA methods~\cite{Hammoudeh2022TrainingDI}. Unrolling the training process results in a computational graph which can be traced backwards to compute attribution. Hence, these methods also consider the training procedure, but differentiate through it to approximate the change in model parameters with leave-one-out retraining (cf. equation~\ref{eq: loo})~\cite{hara2019data}.

Since differentiating through each training step is computationally expensive, methods often group several steps in training segments and base their approximation on a sum of influence function-like computations per segment. Similar to influence functions, \citet{hara2019data} leverage the implicit function theorem to compute the Hessian vector product, and \citet{bae2024training} make use of EKFAC approximations of the Hessian~\cite{grosse2023studying} for the influence-like computation. \citet{ilyas2025magic} propose a different approach using metagradients~\cite{engstrom2025optimizing} to explicitly compute the influence function for single-model training data attribution, fixing sources of training process stochasticity such as batch ordering and model initialisation. While this approach achieves near-optimal performance, it is very costly. 

As unrolling-based methods, like influence functions, aim at capturing the effect of leave-one-out retraining, they provide methods to consider deep model training specificities (e.g., learning rate schedules) and generally target the attribution of model behaviour measured in \textit{loss} on a \textit{single} test sample to a \textit{single} training sample when it is \textit{removed} from training.

\subsection{Group Attribution Methods}
\label{sec:group_if}

TDA methods generally address the individual attribution of the model behaviour $f_\theta(x_{\mathrm{test}})$ to a training sample $z_{\mathrm{train}}$. Another line of work studies methods for quantifying the \emph{group} attribution of $f_\theta(x_{\mathrm{test}})$ to a set of training samples $Z_{\mathrm{train}}$. The TDA objective is analogous to individual attribution, and is generally quantified by the loss difference after leave-\textbf{\textit{some}}-out retraining: 

\begin{equation}
    \label{eq:lso}
    \tau(Z_{\mathrm{train}}, z_{\mathrm{test}};f_\theta) := \mathcal{L}\left(f_{\theta_{\setminus Z_{\mathrm{train}}}}(x_{\mathrm{test}}), y_\mathrm{test} \right) - \mathcal{L}\left(f_\theta(x_{\mathrm{test}}), y_\mathrm{test} \right).
\end{equation}

Like individual attribution, computing equation~\ref{eq:lso} directly is computationally prohibitive when the objective is to identify the group of training samples with the largest attribution. Hence, group TDA work studies methods to approximate this quantity efficiently. 

\paragraph{Group attribution assuming linearity}
An approach that limits the computational cost makes the core assumption that group attribution is linear: In other words, the attribution score of a group of training samples $Z_{\mathrm{train}}$ is defined as the sum of the individual attribution scores of the training samples $z\in Z_{\mathrm{train}}$. \citet{koh2019accuracy} found that interaction effects between training samples of a group can be neglected, and that the linearity assumption is sufficiently accurate for group TDA estimation. As a result, all individual TDA methods, like influence functions, can also be used to estimate group TDA. Moreover, the common evaluation metric \textit{linear datamodeling score}~\cite{park2023trak} used to evaluate the performance of TDA methods is based on the assumption of linearity. 

\paragraph{Group attribution considering interaction effects}
Group TDA methods that assume linearity have the same computational cost as individual TDA methods, but ignore any potential interaction effects between training samples of a group. While \citet{koh2019accuracy} found the interaction effects to be minimal and therefore negligible, a different work by \citet{basu2020second} improves the accuracy of influence functions in capturing equation~\ref{eq:lso} by extending influence functions with a second-order term to capture interaction effects. Shapley value-based approaches like Data Shapley~\cite{ghorbani2019data, lin2024efficient} also consider interaction effects by default. \citet{hu2024most} propose an adaptive algorithm that iterates over the training dataset and gradually identifies the most influential subset for a given model behaviour.

%% file: main.bbl

\begin{thebibliography}{96}


\ifx \showCODEN    \undefined \def \showCODEN     #1{\unskip}     \fi
\ifx \showISBNx    \undefined \def \showISBNx     #1{\unskip}     \fi
\ifx \showISBNxiii \undefined \def \showISBNxiii  #1{\unskip}     \fi
\ifx \showISSN     \undefined \def \showISSN      #1{\unskip}     \fi
\ifx \showLCCN     \undefined \def \showLCCN      #1{\unskip}     \fi
\ifx \shownote     \undefined \def \shownote      #1{#1}          \fi
\ifx \showarticletitle \undefined \def \showarticletitle #1{#1}   \fi
\ifx \showURL      \undefined \def \showURL       {\relax}        \fi
\providecommand\bibfield[2]{#2}
\providecommand\bibinfo[2]{#2}
\providecommand\natexlab[1]{#1}
\providecommand\showeprint[2][]{arXiv:#2}

\bibitem[Adebayo et~al\mbox{.}(2018)]%
        {adebayo2018sanity}
\bibfield{author}{\bibinfo{person}{Julius Adebayo}, \bibinfo{person}{Justin Gilmer}, \bibinfo{person}{Michael Muelly}, \bibinfo{person}{Ian Goodfellow}, \bibinfo{person}{Moritz Hardt}, {and} \bibinfo{person}{Been Kim}.} \bibinfo{year}{2018}\natexlab{}.
\newblock \showarticletitle{Sanity checks for saliency maps}.
\newblock \bibinfo{journal}{\emph{Advances in neural information processing systems}}  \bibinfo{volume}{31} (\bibinfo{year}{2018}).
\newblock


\bibitem[Agarwal et~al\mbox{.}(2017)]%
        {agarwal2017second}
\bibfield{author}{\bibinfo{person}{Naman Agarwal}, \bibinfo{person}{Brian Bullins}, {and} \bibinfo{person}{Elad Hazan}.} \bibinfo{year}{2017}\natexlab{}.
\newblock \showarticletitle{Second-order stochastic optimization for machine learning in linear time}.
\newblock \bibinfo{journal}{\emph{Journal of Machine Learning Research}} \bibinfo{volume}{18}, \bibinfo{number}{116} (\bibinfo{year}{2017}), \bibinfo{pages}{1--40}.
\newblock


\bibitem[Aky{\"u}rek et~al\mbox{.}(2022)]%
        {akyurek2022towards}
\bibfield{author}{\bibinfo{person}{Ekin Aky{\"u}rek}, \bibinfo{person}{Tolga Bolukbasi}, \bibinfo{person}{Frederick Liu}, \bibinfo{person}{Binbin Xiong}, \bibinfo{person}{Ian Tenney}, \bibinfo{person}{Jacob Andreas}, {and} \bibinfo{person}{Kelvin Guu}.} \bibinfo{year}{2022}\natexlab{}.
\newblock \showarticletitle{Towards Tracing Knowledge in Language Models Back to the Training Data}.
\newblock \bibinfo{journal}{\emph{Findings of the Association for Computational Linguistics: EMNLP 2022}} (\bibinfo{year}{2022}), \bibinfo{pages}{2429--2446}.
\newblock


\bibitem[Alsakran et~al\mbox{.}(2014)]%
        {alsakran2014using}
\bibfield{author}{\bibinfo{person}{Jamal Alsakran}, \bibinfo{person}{Xiaoke Huang}, \bibinfo{person}{Ye Zhao}, \bibinfo{person}{Jing Yang}, {and} \bibinfo{person}{Karl Fast}.} \bibinfo{year}{2014}\natexlab{}.
\newblock \showarticletitle{Using entropy-related measures in categorical data visualization}. In \bibinfo{booktitle}{\emph{2014 IEEE Pacific Visualization Symposium}}. IEEE, \bibinfo{pages}{81--88}.
\newblock


\bibitem[Arnoldi(1951)]%
        {arnoldi1951principle}
\bibfield{author}{\bibinfo{person}{Walter~Edwin Arnoldi}.} \bibinfo{year}{1951}\natexlab{}.
\newblock \showarticletitle{The principle of minimized iterations in the solution of the matrix eigenvalue problem}.
\newblock \bibinfo{journal}{\emph{Quarterly of applied mathematics}} \bibinfo{volume}{9}, \bibinfo{number}{1} (\bibinfo{year}{1951}), \bibinfo{pages}{17--29}.
\newblock


\bibitem[Bae et~al\mbox{.}(2024)]%
        {bae2024training}
\bibfield{author}{\bibinfo{person}{Juhan Bae}, \bibinfo{person}{Wu Lin}, \bibinfo{person}{Jonathan Lorraine}, {and} \bibinfo{person}{Roger Grosse}.} \bibinfo{year}{2024}\natexlab{}.
\newblock \showarticletitle{Training Data Attribution via Approximate Unrolled Differentation}.
\newblock \bibinfo{journal}{\emph{arXiv preprint arXiv:2405.12186}} (\bibinfo{year}{2024}).
\newblock


\bibitem[Bae et~al\mbox{.}(2022)]%
        {bae_if_2022}
\bibfield{author}{\bibinfo{person}{Juhan Bae}, \bibinfo{person}{Nathan Ng}, \bibinfo{person}{Alston Lo}, \bibinfo{person}{Marzyeh Ghassemi}, {and} \bibinfo{person}{Roger~B Grosse}.} \bibinfo{year}{2022}\natexlab{}.
\newblock \showarticletitle{If Influence Functions are the Answer, Then What is the Question?}. In \bibinfo{booktitle}{\emph{Advances in Neural Information Processing Systems}}, \bibfield{editor}{\bibinfo{person}{S.~Koyejo}, \bibinfo{person}{S.~Mohamed}, \bibinfo{person}{A.~Agarwal}, \bibinfo{person}{D.~Belgrave}, \bibinfo{person}{K.~Cho}, {and} \bibinfo{person}{A.~Oh}} (Eds.), Vol.~\bibinfo{volume}{35}. \bibinfo{publisher}{Curran Associates, Inc.}, \bibinfo{pages}{17953--17967}.
\newblock
\urldef\tempurl%
\url{https://proceedings.neurips.cc/paper_files/paper/2022/file/7234e0c36fdbcb23e7bd56b68838999b-Paper-Conference.pdf}
\showURL{%
\tempurl}


\bibitem[Bareeva et~al\mbox{.}(2024)]%
        {bareeva2024quandainterpretabilitytoolkittraining}
\bibfield{author}{\bibinfo{person}{Dilyara Bareeva}, \bibinfo{person}{Galip Ümit Yolcu}, \bibinfo{person}{Anna Hedström}, \bibinfo{person}{Niklas Schmolenski}, \bibinfo{person}{Thomas Wiegand}, \bibinfo{person}{Wojciech Samek}, {and} \bibinfo{person}{Sebastian Lapuschkin}.} \bibinfo{year}{2024}\natexlab{}.
\newblock \bibinfo{title}{Quanda: An Interpretability Toolkit for Training Data Attribution Evaluation and Beyond}.
\newblock
\showeprint[arxiv]{2410.07158}~[cs.LG]
\urldef\tempurl%
\url{https://arxiv.org/abs/2410.07158}
\showURL{%
\tempurl}


\bibitem[Basu et~al\mbox{.}(2021)]%
        {basu_influence_2021}
\bibfield{author}{\bibinfo{person}{Samyadeep Basu}, \bibinfo{person}{Phil Pope}, {and} \bibinfo{person}{Soheil Feizi}.} \bibinfo{year}{2021}\natexlab{}.
\newblock \showarticletitle{Influence Functions in Deep Learning Are Fragile}. In \bibinfo{booktitle}{\emph{International Conference on Learning Representations}}.
\newblock
\urldef\tempurl%
\url{https://openreview.net/forum?id=xHKVVHGDOEk}
\showURL{%
\tempurl}


\bibitem[Basu et~al\mbox{.}(2020)]%
        {basu2020second}
\bibfield{author}{\bibinfo{person}{Samyadeep Basu}, \bibinfo{person}{Xuchen You}, {and} \bibinfo{person}{Soheil Feizi}.} \bibinfo{year}{2020}\natexlab{}.
\newblock \showarticletitle{On second-order group influence functions for black-box predictions}. In \bibinfo{booktitle}{\emph{International Conference on Machine Learning}}. PMLR, \bibinfo{pages}{715--724}.
\newblock


\bibitem[Bilodeau et~al\mbox{.}(2024)]%
        {bilodeau2024impossibility}
\bibfield{author}{\bibinfo{person}{Blair Bilodeau}, \bibinfo{person}{Natasha Jaques}, \bibinfo{person}{Pang~Wei Koh}, {and} \bibinfo{person}{Been Kim}.} \bibinfo{year}{2024}\natexlab{}.
\newblock \showarticletitle{Impossibility theorems for feature attribution}.
\newblock \bibinfo{journal}{\emph{Proceedings of the National Academy of Sciences}} \bibinfo{volume}{121}, \bibinfo{number}{2} (\bibinfo{year}{2024}), \bibinfo{pages}{e2304406120}.
\newblock


\bibitem[Braun and Clarke(2012)]%
        {braun2012thematic}
\bibfield{author}{\bibinfo{person}{Virginia Braun} {and} \bibinfo{person}{Victoria Clarke}.} \bibinfo{year}{2012}\natexlab{}.
\newblock \bibinfo{booktitle}{\emph{Thematic analysis.}}
\newblock \bibinfo{publisher}{American Psychological Association}.
\newblock


\bibitem[Brennen(2020)]%
        {brennen2020what}
\bibfield{author}{\bibinfo{person}{Andrea Brennen}.} \bibinfo{year}{2020}\natexlab{}.
\newblock \showarticletitle{What Do People Really Want When They Say They Want "Explainable AI?" We Asked 60 Stakeholders.}. In \bibinfo{booktitle}{\emph{Extended Abstracts of the 2020 CHI Conference on Human Factors in Computing Systems}} (Honolulu, HI, USA) \emph{(\bibinfo{series}{CHI EA '20})}. \bibinfo{publisher}{Association for Computing Machinery}, \bibinfo{address}{New York, NY, USA}, \bibinfo{pages}{1–7}.
\newblock
\showISBNx{9781450368193}
\href{https://doi.org/10.1145/3334480.3383047}{doi:\nolinkurl{10.1145/3334480.3383047}}


\bibitem[Brunet et~al\mbox{.}(2019)]%
        {brunet2019understanding}
\bibfield{author}{\bibinfo{person}{Marc-Etienne Brunet}, \bibinfo{person}{Colleen Alkalay-Houlihan}, \bibinfo{person}{Ashton Anderson}, {and} \bibinfo{person}{Richard Zemel}.} \bibinfo{year}{2019}\natexlab{}.
\newblock \showarticletitle{Understanding the origins of bias in word embeddings}. In \bibinfo{booktitle}{\emph{International conference on machine learning}}. PMLR, \bibinfo{pages}{803--811}.
\newblock


\bibitem[Cabrera et~al\mbox{.}(2023)]%
        {cabrera2023what}
\bibfield{author}{\bibinfo{person}{\'{A}ngel~Alexander Cabrera}, \bibinfo{person}{Marco Tulio~Ribeiro}, \bibinfo{person}{Bongshin Lee}, \bibinfo{person}{Robert Deline}, \bibinfo{person}{Adam Perer}, {and} \bibinfo{person}{Steven~M. Drucker}.} \bibinfo{year}{2023}\natexlab{}.
\newblock \showarticletitle{What Did My AI Learn? How Data Scientists Make Sense of Model Behavior}.
\newblock \bibinfo{journal}{\emph{ACM Trans. Comput.-Hum. Interact.}} \bibinfo{volume}{30}, \bibinfo{number}{1}, Article \bibinfo{articleno}{1} (\bibinfo{date}{March} \bibinfo{year}{2023}), \bibinfo{numpages}{27}~pages.
\newblock
\showISSN{1073-0516}
\href{https://doi.org/10.1145/3542921}{doi:\nolinkurl{10.1145/3542921}}


\bibitem[Carroll(1995)]%
        {carroll1995sbd}
\bibfield{editor}{\bibinfo{person}{John~M. Carroll}} (Ed.). \bibinfo{year}{1995}\natexlab{}.
\newblock \bibinfo{booktitle}{\emph{Scenario-based design: envisioning work and technology in system development}}.
\newblock \bibinfo{publisher}{John Wiley \& Sons, Inc.}, \bibinfo{address}{USA}.
\newblock
\showISBNx{0471076597}


\bibitem[Charpiat et~al\mbox{.}(2019)]%
        {charpiat_input_2021}
\bibfield{author}{\bibinfo{person}{Guillaume Charpiat}, \bibinfo{person}{Nicolas Girard}, \bibinfo{person}{Loris Felardos}, {and} \bibinfo{person}{Yuliya Tarabalka}.} \bibinfo{year}{2019}\natexlab{}.
\newblock \showarticletitle{Input Similarity from the Neural Network Perspective}. In \bibinfo{booktitle}{\emph{Advances in Neural Information Processing Systems}}, \bibfield{editor}{\bibinfo{person}{H.~Wallach}, \bibinfo{person}{H.~Larochelle}, \bibinfo{person}{A.~Beygelzimer}, \bibinfo{person}{F.~d\textquotesingle Alch\'{e}-Buc}, \bibinfo{person}{E.~Fox}, {and} \bibinfo{person}{R.~Garnett}} (Eds.), Vol.~\bibinfo{volume}{32}. \bibinfo{publisher}{Curran Associates, Inc.}
\newblock
\urldef\tempurl%
\url{https://proceedings.neurips.cc/paper_files/paper/2019/file/c61f571dbd2fb949d3fe5ae1608dd48b-Paper.pdf}
\showURL{%
\tempurl}


\bibitem[Choe et~al\mbox{.}(2024)]%
        {choe2024your}
\bibfield{author}{\bibinfo{person}{Sang~Keun Choe}, \bibinfo{person}{Hwijeen Ahn}, \bibinfo{person}{Juhan Bae}, \bibinfo{person}{Kewen Zhao}, \bibinfo{person}{Minsoo Kang}, \bibinfo{person}{Youngseog Chung}, \bibinfo{person}{Adithya Pratapa}, \bibinfo{person}{Willie Neiswanger}, \bibinfo{person}{Emma Strubell}, \bibinfo{person}{Teruko Mitamura}, {et~al\mbox{.}}} \bibinfo{year}{2024}\natexlab{}.
\newblock \showarticletitle{What is Your Data Worth to GPT? LLM-Scale Data Valuation with Influence Functions}.
\newblock \bibinfo{journal}{\emph{arXiv preprint arXiv:2405.13954}} (\bibinfo{year}{2024}).
\newblock


\bibitem[Deng et~al\mbox{.}(2024)]%
        {deng2024texttt}
\bibfield{author}{\bibinfo{person}{Junwei Deng}, \bibinfo{person}{Ting-Wei Li}, \bibinfo{person}{Shiyuan Zhang}, \bibinfo{person}{Shixuan Liu}, \bibinfo{person}{Yijun Pan}, \bibinfo{person}{Hao Huang}, \bibinfo{person}{Xinhe Wang}, \bibinfo{person}{Pingbang Hu}, \bibinfo{person}{Xingjian Zhang}, {and} \bibinfo{person}{Jiaqi Ma}.} \bibinfo{year}{2024}\natexlab{}.
\newblock \showarticletitle{dattri: A Library for Efficient Data Attribution}.
\newblock \bibinfo{journal}{\emph{Advances in Neural Information Processing Systems}}  \bibinfo{volume}{37} (\bibinfo{year}{2024}), \bibinfo{pages}{136763--136781}.
\newblock


\bibitem[Doshi-Velez and Kim(2017)]%
        {doshi2017towards}
\bibfield{author}{\bibinfo{person}{Finale Doshi-Velez} {and} \bibinfo{person}{Been Kim}.} \bibinfo{year}{2017}\natexlab{}.
\newblock \showarticletitle{Towards a rigorous science of interpretable machine learning}.
\newblock \bibinfo{journal}{\emph{arXiv preprint arXiv:1702.08608}} (\bibinfo{year}{2017}).
\newblock


\bibitem[Dwivedi et~al\mbox{.}(2023)]%
        {dwivedi2023explainable}
\bibfield{author}{\bibinfo{person}{Rudresh Dwivedi}, \bibinfo{person}{Devam Dave}, \bibinfo{person}{Het Naik}, \bibinfo{person}{Smiti Singhal}, \bibinfo{person}{Rana Omer}, \bibinfo{person}{Pankesh Patel}, \bibinfo{person}{Bin Qian}, \bibinfo{person}{Zhenyu Wen}, \bibinfo{person}{Tejal Shah}, \bibinfo{person}{Graham Morgan}, {et~al\mbox{.}}} \bibinfo{year}{2023}\natexlab{}.
\newblock \showarticletitle{Explainable AI (XAI): Core ideas, techniques, and solutions}.
\newblock \bibinfo{journal}{\emph{Comput. Surveys}} \bibinfo{volume}{55}, \bibinfo{number}{9} (\bibinfo{year}{2023}), \bibinfo{pages}{1--33}.
\newblock


\bibitem[Ehsan et~al\mbox{.}(2021)]%
        {ehsan2021expanding}
\bibfield{author}{\bibinfo{person}{Upol Ehsan}, \bibinfo{person}{Q~Vera Liao}, \bibinfo{person}{Michael Muller}, \bibinfo{person}{Mark~O Riedl}, {and} \bibinfo{person}{Justin~D Weisz}.} \bibinfo{year}{2021}\natexlab{}.
\newblock \showarticletitle{Expanding explainability: Towards social transparency in ai systems}. In \bibinfo{booktitle}{\emph{Proceedings of the 2021 CHI conference on human factors in computing systems}}. \bibinfo{pages}{1--19}.
\newblock


\bibitem[Ehsan et~al\mbox{.}(2024)]%
        {ehsan2024who}
\bibfield{author}{\bibinfo{person}{Upol Ehsan}, \bibinfo{person}{Samir Passi}, \bibinfo{person}{Q.~Vera Liao}, \bibinfo{person}{Larry Chan}, \bibinfo{person}{I-Hsiang Lee}, \bibinfo{person}{Michael Muller}, {and} \bibinfo{person}{Mark~O Riedl}.} \bibinfo{year}{2024}\natexlab{}.
\newblock \showarticletitle{The Who in XAI: How AI Background Shapes Perceptions of AI Explanations}. In \bibinfo{booktitle}{\emph{Proceedings of the CHI Conference on Human Factors in Computing Systems}} (Honolulu, HI, USA) \emph{(\bibinfo{series}{CHI '24})}. \bibinfo{publisher}{Association for Computing Machinery}, \bibinfo{address}{New York, NY, USA}, Article \bibinfo{articleno}{316}, \bibinfo{numpages}{32}~pages.
\newblock
\showISBNx{9798400703300}
\href{https://doi.org/10.1145/3613904.3642474}{doi:\nolinkurl{10.1145/3613904.3642474}}


\bibitem[Ehsan and Riedl(2020)]%
        {ehsan2020hcxai}
\bibfield{author}{\bibinfo{person}{Upol Ehsan} {and} \bibinfo{person}{Mark~O. Riedl}.} \bibinfo{year}{2020}\natexlab{}.
\newblock \showarticletitle{Human-Centered Explainable AI: Towards a Reflective Sociotechnical Approach}. In \bibinfo{booktitle}{\emph{HCI International 2020 - Late Breaking Papers: Multimodality and Intelligence}}, \bibfield{editor}{\bibinfo{person}{Constantine Stephanidis}, \bibinfo{person}{Masaaki Kurosu}, \bibinfo{person}{Helmut Degen}, {and} \bibinfo{person}{Lauren Reinerman-Jones}} (Eds.). \bibinfo{publisher}{Springer International Publishing}, \bibinfo{address}{Cham}, \bibinfo{pages}{449--466}.
\newblock
\showISBNx{978-3-030-60117-1}


\bibitem[Engstrom et~al\mbox{.}(2025)]%
        {engstrom2025optimizing}
\bibfield{author}{\bibinfo{person}{Logan Engstrom}, \bibinfo{person}{Andrew Ilyas}, \bibinfo{person}{Benjamin Chen}, \bibinfo{person}{Axel Feldmann}, \bibinfo{person}{William Moses}, {and} \bibinfo{person}{Aleksander Madry}.} \bibinfo{year}{2025}\natexlab{}.
\newblock \showarticletitle{Optimizing ml training with metagradient descent}.
\newblock \bibinfo{journal}{\emph{arXiv preprint arXiv:2503.13751}} (\bibinfo{year}{2025}).
\newblock


\bibitem[Epifano et~al\mbox{.}(2023)]%
        {epifano_revisiting_2023}
\bibfield{author}{\bibinfo{person}{Jacob Epifano}, \bibinfo{person}{Ravichandran Ramachandran}, \bibinfo{person}{Aaron~J. Masino}, {and} \bibinfo{person}{Ghulam Rasool}.} \bibinfo{year}{2023}\natexlab{}.
\newblock \showarticletitle{Revisiting the Fragility of Influence Functions}.
\newblock \bibinfo{journal}{\emph{Neural networks : the official journal of the International Neural Network Society}}  \bibinfo{volume}{162} (\bibinfo{year}{2023}), \bibinfo{pages}{581--588}.
\newblock


\bibitem[Feldman and Zhang(2020)]%
        {feldman2020neural}
\bibfield{author}{\bibinfo{person}{Vitaly Feldman} {and} \bibinfo{person}{Chiyuan Zhang}.} \bibinfo{year}{2020}\natexlab{}.
\newblock \showarticletitle{What neural networks memorize and why: Discovering the long tail via influence estimation}.
\newblock \bibinfo{journal}{\emph{Advances in Neural Information Processing Systems}}  \bibinfo{volume}{33} (\bibinfo{year}{2020}), \bibinfo{pages}{2881--2891}.
\newblock


\bibitem[Fok and Weld(2023)]%
        {fok2023search}
\bibfield{author}{\bibinfo{person}{Raymond Fok} {and} \bibinfo{person}{Daniel~S Weld}.} \bibinfo{year}{2023}\natexlab{}.
\newblock \showarticletitle{In search of verifiability: Explanations rarely enable complementary performance in AI-advised decision making}.
\newblock \bibinfo{journal}{\emph{AI Magazine}} (\bibinfo{year}{2023}).
\newblock


\bibitem[Ghorbani and Zou(2019)]%
        {ghorbani2019data}
\bibfield{author}{\bibinfo{person}{Amirata Ghorbani} {and} \bibinfo{person}{James Zou}.} \bibinfo{year}{2019}\natexlab{}.
\newblock \showarticletitle{Data shapley: Equitable valuation of data for machine learning}. In \bibinfo{booktitle}{\emph{International conference on machine learning}}. PMLR, \bibinfo{pages}{2242--2251}.
\newblock


\bibitem[Grosse et~al\mbox{.}(2023)]%
        {grosse2023studying}
\bibfield{author}{\bibinfo{person}{Roger Grosse}, \bibinfo{person}{Juhan Bae}, \bibinfo{person}{Cem Anil}, \bibinfo{person}{Nelson Elhage}, \bibinfo{person}{Alex Tamkin}, \bibinfo{person}{Amirhossein Tajdini}, \bibinfo{person}{Benoit Steiner}, \bibinfo{person}{Dustin Li}, \bibinfo{person}{Esin Durmus}, \bibinfo{person}{Ethan Perez}, {et~al\mbox{.}}} \bibinfo{year}{2023}\natexlab{}.
\newblock \showarticletitle{Studying large language model generalization with influence functions}.
\newblock \bibinfo{journal}{\emph{arXiv preprint arXiv:2308.03296}} (\bibinfo{year}{2023}).
\newblock


\bibitem[Guest et~al\mbox{.}(2023)]%
        {guest_collecting_2023}
\bibfield{author}{\bibinfo{person}{Greg Guest}, \bibinfo{person}{Emily~E. Namey}, {and} \bibinfo{person}{Marilyn~L. Mitchell}.} \bibinfo{year}{2023}\natexlab{}.
\newblock \showarticletitle{Collecting {Qualitative} {Data}: {A} {Field} {Manual} for {Applied} {Research}}.
\newblock \bibinfo{publisher}{SAGE Publications, Ltd}, \bibinfo{address}{55 City Road}.
\newblock
\href{https://doi.org/10.4135/9781506374680}{doi:\nolinkurl{10.4135/9781506374680}}


\bibitem[Guidotti et~al\mbox{.}(2018)]%
        {guidotti2018}
\bibfield{author}{\bibinfo{person}{Riccardo Guidotti}, \bibinfo{person}{Anna Monreale}, \bibinfo{person}{Salvatore Ruggieri}, \bibinfo{person}{Franco Turini}, \bibinfo{person}{Fosca Giannotti}, {and} \bibinfo{person}{Dino Pedreschi}.} \bibinfo{year}{2018}\natexlab{}.
\newblock \showarticletitle{A Survey of Methods for Explaining Black Box Models}.
\newblock \bibinfo{journal}{\emph{ACM Comput. Surv.}} \bibinfo{volume}{51}, \bibinfo{number}{5}, Article \bibinfo{articleno}{93} (\bibinfo{year}{2018}), \bibinfo{numpages}{42}~pages.
\newblock
\showISSN{0360-0300}
\href{https://doi.org/10.1145/3236009}{doi:\nolinkurl{10.1145/3236009}}


\bibitem[Guo et~al\mbox{.}(2021)]%
        {guo-etal-2021-fastif}
\bibfield{author}{\bibinfo{person}{Han Guo}, \bibinfo{person}{Nazneen Rajani}, \bibinfo{person}{Peter Hase}, \bibinfo{person}{Mohit Bansal}, {and} \bibinfo{person}{Caiming Xiong}.} \bibinfo{year}{2021}\natexlab{}.
\newblock \showarticletitle{{F}ast{IF}: Scalable Influence Functions for Efficient Model Interpretation and Debugging}. In \bibinfo{booktitle}{\emph{Proceedings of the 2021 Conference on Empirical Methods in Natural Language Processing}}. \bibinfo{publisher}{Association for Computational Linguistics}, \bibinfo{address}{Online and Punta Cana, Dominican Republic}, \bibinfo{pages}{10333--10350}.
\newblock
\href{https://doi.org/10.18653/v1/2021.emnlp-main.808}{doi:\nolinkurl{10.18653/v1/2021.emnlp-main.808}}


\bibitem[Hadash et~al\mbox{.}(2022)]%
        {hadash2022improving}
\bibfield{author}{\bibinfo{person}{Sophia Hadash}, \bibinfo{person}{Martijn~C. Willemsen}, \bibinfo{person}{Chris Snijders}, {and} \bibinfo{person}{Wijnand~A. IJsselsteijn}.} \bibinfo{year}{2022}\natexlab{}.
\newblock \showarticletitle{Improving understandability of feature contributions in model-agnostic explainable AI tools}. In \bibinfo{booktitle}{\emph{Proceedings of the 2022 CHI Conference on Human Factors in Computing Systems}} (New Orleans, LA, USA) \emph{(\bibinfo{series}{CHI '22})}. \bibinfo{publisher}{Association for Computing Machinery}, \bibinfo{address}{New York, NY, USA}, Article \bibinfo{articleno}{487}, \bibinfo{numpages}{9}~pages.
\newblock
\showISBNx{9781450391573}
\href{https://doi.org/10.1145/3491102.3517650}{doi:\nolinkurl{10.1145/3491102.3517650}}


\bibitem[Hammoudeh and Lowd(2024)]%
        {Hammoudeh2022TrainingDI}
\bibfield{author}{\bibinfo{person}{Zayd Hammoudeh} {and} \bibinfo{person}{Daniel Lowd}.} \bibinfo{year}{2024}\natexlab{}.
\newblock \showarticletitle{Training data influence analysis and estimation: A survey}.
\newblock \bibinfo{journal}{\emph{Machine Learning}} \bibinfo{volume}{113}, \bibinfo{number}{5} (\bibinfo{year}{2024}), \bibinfo{pages}{2351--2403}.
\newblock


\bibitem[Hampel(1974)]%
        {hampel1974}
\bibfield{author}{\bibinfo{person}{Frank~R. Hampel}.} \bibinfo{year}{1974}\natexlab{}.
\newblock \showarticletitle{The Influence Curve and its Role in Robust Estimation}.
\newblock \bibinfo{journal}{\emph{J. Amer. Statist. Assoc.}} \bibinfo{volume}{69}, \bibinfo{number}{346} (\bibinfo{year}{1974}), \bibinfo{pages}{383--393}.
\newblock
\href{https://doi.org/10.1080/01621459.1974.10482962}{doi:\nolinkurl{10.1080/01621459.1974.10482962}}


\bibitem[Hara et~al\mbox{.}(2019)]%
        {hara2019data}
\bibfield{author}{\bibinfo{person}{Satoshi Hara}, \bibinfo{person}{Atsushi Nitanda}, {and} \bibinfo{person}{Takanori Maehara}.} \bibinfo{year}{2019}\natexlab{}.
\newblock \showarticletitle{Data cleansing for models trained with SGD}.
\newblock \bibinfo{journal}{\emph{Advances in Neural Information Processing Systems}}  \bibinfo{volume}{32} (\bibinfo{year}{2019}).
\newblock


\bibitem[He et~al\mbox{.}(2016)]%
        {he2016deep}
\bibfield{author}{\bibinfo{person}{Kaiming He}, \bibinfo{person}{Xiangyu Zhang}, \bibinfo{person}{Shaoqing Ren}, {and} \bibinfo{person}{Jian Sun}.} \bibinfo{year}{2016}\natexlab{}.
\newblock \showarticletitle{Deep residual learning for image recognition}. In \bibinfo{booktitle}{\emph{Proceedings of the IEEE conference on computer vision and pattern recognition}}. \bibinfo{pages}{770--778}.
\newblock


\bibitem[Hofmann(1994)]%
        {hofmann1994statlog}
\bibfield{author}{\bibinfo{person}{Hans Hofmann}.} \bibinfo{year}{1994}\natexlab{}.
\newblock \bibinfo{title}{{Statlog (German Credit Data)}}.
\newblock \bibinfo{howpublished}{UCI Machine Learning Repository}.
\newblock
\newblock
\shownote{{DOI}: https://doi.org/10.24432/C5NC77}.


\bibitem[Hohman et~al\mbox{.}(2020)]%
        {hohman2020understanding}
\bibfield{author}{\bibinfo{person}{Fred Hohman}, \bibinfo{person}{Kanit Wongsuphasawat}, \bibinfo{person}{Mary~Beth Kery}, {and} \bibinfo{person}{Kayur Patel}.} \bibinfo{year}{2020}\natexlab{}.
\newblock \showarticletitle{Understanding and Visualizing Data Iteration in Machine Learning}. In \bibinfo{booktitle}{\emph{Proceedings of the 2020 CHI Conference on Human Factors in Computing Systems}} (Honolulu, HI, USA) \emph{(\bibinfo{series}{CHI '20})}. \bibinfo{publisher}{Association for Computing Machinery}, \bibinfo{address}{New York, NY, USA}, \bibinfo{pages}{1–13}.
\newblock
\showISBNx{9781450367080}
\href{https://doi.org/10.1145/3313831.3376177}{doi:\nolinkurl{10.1145/3313831.3376177}}


\bibitem[Hong et~al\mbox{.}(2020)]%
        {hong2020human}
\bibfield{author}{\bibinfo{person}{Sungsoo~Ray Hong}, \bibinfo{person}{Jessica Hullman}, {and} \bibinfo{person}{Enrico Bertini}.} \bibinfo{year}{2020}\natexlab{}.
\newblock \showarticletitle{Human Factors in Model Interpretability: Industry Practices, Challenges, and Needs}.
\newblock \bibinfo{journal}{\emph{Proc. ACM Hum.-Comput. Interact.}} \bibinfo{volume}{4}, \bibinfo{number}{CSCW1}, Article \bibinfo{articleno}{68} (\bibinfo{date}{May} \bibinfo{year}{2020}), \bibinfo{numpages}{26}~pages.
\newblock
\href{https://doi.org/10.1145/3392878}{doi:\nolinkurl{10.1145/3392878}}


\bibitem[Hu et~al\mbox{.}(2024)]%
        {hu2024most}
\bibfield{author}{\bibinfo{person}{Yuzheng Hu}, \bibinfo{person}{Pingbang Hu}, \bibinfo{person}{Han Zhao}, {and} \bibinfo{person}{Jiaqi Ma}.} \bibinfo{year}{2024}\natexlab{}.
\newblock \showarticletitle{Most Influential Subset Selection: Challenges, Promises, and Beyond}. In \bibinfo{booktitle}{\emph{Advances in Neural Information Processing Systems}}, \bibfield{editor}{\bibinfo{person}{A.~Globerson}, \bibinfo{person}{L.~Mackey}, \bibinfo{person}{D.~Belgrave}, \bibinfo{person}{A.~Fan}, \bibinfo{person}{U.~Paquet}, \bibinfo{person}{J.~Tomczak}, {and} \bibinfo{person}{C.~Zhang}} (Eds.), Vol.~\bibinfo{volume}{37}. \bibinfo{publisher}{Curran Associates, Inc.}, \bibinfo{pages}{119778--119810}.
\newblock
\urldef\tempurl%
\url{https://proceedings.neurips.cc/paper_files/paper/2024/file/d8684e49752e06ac5e4b554b60ad212a-Paper-Conference.pdf}
\showURL{%
\tempurl}


\bibitem[Ilyas and Engstrom(2025)]%
        {ilyas2025magic}
\bibfield{author}{\bibinfo{person}{Andrew Ilyas} {and} \bibinfo{person}{Logan Engstrom}.} \bibinfo{year}{2025}\natexlab{}.
\newblock \showarticletitle{MAGIC: Near-Optimal Data Attribution for Deep Learning}.
\newblock \bibinfo{journal}{\emph{arXiv preprint arXiv:2504.16430}} (\bibinfo{year}{2025}).
\newblock


\bibitem[Ilyas et~al\mbox{.}(2022)]%
        {ilyas2022datamodels}
\bibfield{author}{\bibinfo{person}{Andrew Ilyas}, \bibinfo{person}{Sung~Min Park}, \bibinfo{person}{Logan Engstrom}, \bibinfo{person}{Guillaume Leclerc}, {and} \bibinfo{person}{Aleksander Madry}.} \bibinfo{year}{2022}\natexlab{}.
\newblock \showarticletitle{Datamodels: Predicting Predictions from Training Data}. In \bibinfo{booktitle}{\emph{Proceedings of the 39th International Conference on Machine Learning}}.
\newblock


\bibitem[Jain et~al\mbox{.}(2024)]%
        {jain2024data}
\bibfield{author}{\bibinfo{person}{Saachi Jain}, \bibinfo{person}{Kimia Hamidieh}, \bibinfo{person}{Kristian Georgiev}, \bibinfo{person}{Andrew Ilyas}, \bibinfo{person}{Marzyeh Ghassemi}, {and} \bibinfo{person}{Aleksander Madry}.} \bibinfo{year}{2024}\natexlab{}.
\newblock \showarticletitle{Data Debiasing with Datamodels (D3M): Improving Subgroup Robustness via Data Selection}.
\newblock \bibinfo{journal}{\emph{arXiv preprint arXiv:2406.16846}} (\bibinfo{year}{2024}).
\newblock


\bibitem[Jin et~al\mbox{.}(2023)]%
        {jin2023invisible}
\bibfield{author}{\bibinfo{person}{Weina Jin}, \bibinfo{person}{Jianyu Fan}, \bibinfo{person}{Diane Gromala}, \bibinfo{person}{Philippe Pasquier}, {and} \bibinfo{person}{Ghassan Hamarneh}.} \bibinfo{year}{2023}\natexlab{}.
\newblock \showarticletitle{Invisible users: Uncovering end-users' requirements for explainable ai via explanation forms and goals}.
\newblock \bibinfo{journal}{\emph{arXiv preprint arXiv:2302.06609}} (\bibinfo{year}{2023}).
\newblock


\bibitem[Jin et~al\mbox{.}(2024)]%
        {DCAISurvey2024}
\bibfield{author}{\bibinfo{person}{Wei Jin}, \bibinfo{person}{Haohan Wang}, \bibinfo{person}{Daochen Zha}, \bibinfo{person}{Qiaoyu Tan}, \bibinfo{person}{Yao Ma}, \bibinfo{person}{Sharon Li}, {and} \bibinfo{person}{Su-In Lee}.} \bibinfo{year}{2024}\natexlab{}.
\newblock \showarticletitle{DCAI: Data-centric Artificial Intelligence}. In \bibinfo{booktitle}{\emph{Companion Proceedings of the ACM Web Conference 2024}} (Singapore, Singapore) \emph{(\bibinfo{series}{WWW '24})}. \bibinfo{publisher}{Association for Computing Machinery}, \bibinfo{address}{New York, NY, USA}, \bibinfo{pages}{1482–1485}.
\newblock
\showISBNx{9798400701726}
\href{https://doi.org/10.1145/3589335.3641297}{doi:\nolinkurl{10.1145/3589335.3641297}}


\bibitem[Juneja et~al\mbox{.}(2024)]%
        {juneja2024dissecting}
\bibfield{author}{\bibinfo{person}{Prerna Juneja}, \bibinfo{person}{Wenjuan Zhang}, \bibinfo{person}{Alison~Marie Smith-Renner}, \bibinfo{person}{Hemank Lamba}, \bibinfo{person}{Joel Tetreault}, {and} \bibinfo{person}{Alex Jaimes}.} \bibinfo{year}{2024}\natexlab{}.
\newblock \showarticletitle{Dissecting users' needs for search result explanations}. In \bibinfo{booktitle}{\emph{Proceedings of the CHI Conference on Human Factors in Computing Systems}} (Honolulu, HI, USA) \emph{(\bibinfo{series}{CHI '24})}. \bibinfo{publisher}{Association for Computing Machinery}, \bibinfo{address}{New York, NY, USA}, Article \bibinfo{articleno}{841}, \bibinfo{numpages}{17}~pages.
\newblock
\showISBNx{9798400703300}
\href{https://doi.org/10.1145/3613904.3642059}{doi:\nolinkurl{10.1145/3613904.3642059}}


\bibitem[Kaur et~al\mbox{.}(2020)]%
        {kaur2020interpreting}
\bibfield{author}{\bibinfo{person}{Harmanpreet Kaur}, \bibinfo{person}{Harsha Nori}, \bibinfo{person}{Samuel Jenkins}, \bibinfo{person}{Rich Caruana}, \bibinfo{person}{Hanna Wallach}, {and} \bibinfo{person}{Jennifer Wortman~Vaughan}.} \bibinfo{year}{2020}\natexlab{}.
\newblock \showarticletitle{Interpreting Interpretability: Understanding Data Scientists' Use of Interpretability Tools for Machine Learning}. In \bibinfo{booktitle}{\emph{Proceedings of the 2020 CHI Conference on Human Factors in Computing Systems}} (Honolulu, HI, USA) \emph{(\bibinfo{series}{CHI '20})}. \bibinfo{publisher}{Association for Computing Machinery}, \bibinfo{address}{New York, NY, USA}, \bibinfo{pages}{1–14}.
\newblock
\showISBNx{9781450367080}
\href{https://doi.org/10.1145/3313831.3376219}{doi:\nolinkurl{10.1145/3313831.3376219}}


\bibitem[Kim et~al\mbox{.}(2022)]%
        {kim2022hive}
\bibfield{author}{\bibinfo{person}{Sunnie~SY Kim}, \bibinfo{person}{Nicole Meister}, \bibinfo{person}{Vikram~V Ramaswamy}, \bibinfo{person}{Ruth Fong}, {and} \bibinfo{person}{Olga Russakovsky}.} \bibinfo{year}{2022}\natexlab{}.
\newblock \showarticletitle{HIVE: Evaluating the human interpretability of visual explanations}. In \bibinfo{booktitle}{\emph{European Conference on Computer Vision}}. Springer, \bibinfo{pages}{280--298}.
\newblock


\bibitem[Kim et~al\mbox{.}(2023)]%
        {kim2023help}
\bibfield{author}{\bibinfo{person}{Sunnie S.~Y. Kim}, \bibinfo{person}{Elizabeth~Anne Watkins}, \bibinfo{person}{Olga Russakovsky}, \bibinfo{person}{Ruth Fong}, {and} \bibinfo{person}{Andr\'{e}s Monroy-Hern\'{a}ndez}.} \bibinfo{year}{2023}\natexlab{}.
\newblock \showarticletitle{"Help Me Help the AI": Understanding How Explainability Can Support Human-AI Interaction}. In \bibinfo{booktitle}{\emph{Proceedings of the 2023 CHI Conference on Human Factors in Computing Systems}} (Hamburg, Germany) \emph{(\bibinfo{series}{CHI '23})}. \bibinfo{publisher}{Association for Computing Machinery}, \bibinfo{address}{New York, NY, USA}, Article \bibinfo{articleno}{250}, \bibinfo{numpages}{17}~pages.
\newblock
\showISBNx{9781450394215}
\href{https://doi.org/10.1145/3544548.3581001}{doi:\nolinkurl{10.1145/3544548.3581001}}


\bibitem[Koh and Liang(2017)]%
        {kohliang2017}
\bibfield{author}{\bibinfo{person}{Pang~Wei Koh} {and} \bibinfo{person}{Percy Liang}.} \bibinfo{year}{2017}\natexlab{}.
\newblock \showarticletitle{Understanding Black-box Predictions via Influence Functions}. In \bibinfo{booktitle}{\emph{Proceedings of the 34th International Conference on Machine Learning}} \emph{(\bibinfo{series}{Proceedings of Machine Learning Research}, Vol.~\bibinfo{volume}{70})}, \bibfield{editor}{\bibinfo{person}{Doina Precup} {and} \bibinfo{person}{Yee~Whye Teh}} (Eds.). \bibinfo{publisher}{PMLR}, \bibinfo{pages}{1885--1894}.
\newblock
\urldef\tempurl%
\url{https://proceedings.mlr.press/v70/koh17a.html}
\showURL{%
\tempurl}


\bibitem[Koh et~al\mbox{.}(2019)]%
        {koh2019accuracy}
\bibfield{author}{\bibinfo{person}{Pang Wei~W Koh}, \bibinfo{person}{Kai-Siang Ang}, \bibinfo{person}{Hubert Teo}, {and} \bibinfo{person}{Percy~S Liang}.} \bibinfo{year}{2019}\natexlab{}.
\newblock \showarticletitle{On the accuracy of influence functions for measuring group effects}.
\newblock \bibinfo{journal}{\emph{Advances in neural information processing systems}}  \bibinfo{volume}{32} (\bibinfo{year}{2019}).
\newblock


\bibitem[Krippendorff(2011)]%
        {krippendorff2011computing}
\bibfield{author}{\bibinfo{person}{Klaus Krippendorff}.} \bibinfo{year}{2011}\natexlab{}.
\newblock \bibinfo{title}{Computing Krippendorff’s alpha-reliability}.
\newblock


\bibitem[Kwon et~al\mbox{.}(2023)]%
        {kwon2023datainf}
\bibfield{author}{\bibinfo{person}{Yongchan Kwon}, \bibinfo{person}{Eric Wu}, \bibinfo{person}{Kevin Wu}, {and} \bibinfo{person}{James Zou}.} \bibinfo{year}{2023}\natexlab{}.
\newblock \showarticletitle{Datainf: Efficiently estimating data influence in lora-tuned llms and diffusion models}.
\newblock \bibinfo{journal}{\emph{arXiv preprint arXiv:2310.00902}} (\bibinfo{year}{2023}).
\newblock


\bibitem[Lakkaraju et~al\mbox{.}(2022)]%
        {lakkaraju2022rethinking}
\bibfield{author}{\bibinfo{person}{Himabindu Lakkaraju}, \bibinfo{person}{Dylan Slack}, \bibinfo{person}{Yuxin Chen}, \bibinfo{person}{Chenhao Tan}, {and} \bibinfo{person}{Sameer Singh}.} \bibinfo{year}{2022}\natexlab{}.
\newblock \showarticletitle{Rethinking explainability as a dialogue: A practitioner's perspective}.
\newblock \bibinfo{journal}{\emph{arXiv preprint arXiv:2202.01875}} (\bibinfo{year}{2022}).
\newblock


\bibitem[Li et~al\mbox{.}(2024)]%
        {li2024delta}
\bibfield{author}{\bibinfo{person}{Wenjie Li}, \bibinfo{person}{Jiawei Li}, \bibinfo{person}{Christian~Schroeder de Witt}, \bibinfo{person}{Ameya Prabhu}, {and} \bibinfo{person}{Amartya Sanyal}.} \bibinfo{year}{2024}\natexlab{}.
\newblock \showarticletitle{Delta-Influence: Unlearning Poisons via Influence Functions}.
\newblock \bibinfo{journal}{\emph{arXiv preprint arXiv:2411.13731}} (\bibinfo{year}{2024}).
\newblock


\bibitem[Liao et~al\mbox{.}(2020)]%
        {liao2020questionbank}
\bibfield{author}{\bibinfo{person}{Q.~Vera Liao}, \bibinfo{person}{Daniel Gruen}, {and} \bibinfo{person}{Sarah Miller}.} \bibinfo{year}{2020}\natexlab{}.
\newblock \showarticletitle{Questioning the AI: Informing Design Practices for Explainable AI User Experiences}. In \bibinfo{booktitle}{\emph{Proceedings of the 2020 CHI Conference on Human Factors in Computing Systems}} (Honolulu, HI, USA) \emph{(\bibinfo{series}{CHI '20})}. \bibinfo{publisher}{Association for Computing Machinery}, \bibinfo{address}{New York, NY, USA}, \bibinfo{pages}{1–15}.
\newblock
\showISBNx{9781450367080}
\href{https://doi.org/10.1145/3313831.3376590}{doi:\nolinkurl{10.1145/3313831.3376590}}


\bibitem[Lin et~al\mbox{.}(2024)]%
        {lin2024efficient}
\bibfield{author}{\bibinfo{person}{Chris Lin}, \bibinfo{person}{Mingyu Lu}, \bibinfo{person}{Chanwoo Kim}, {and} \bibinfo{person}{Su-In Lee}.} \bibinfo{year}{2024}\natexlab{}.
\newblock \showarticletitle{Efficient Shapley Values for Attributing Global Properties of Diffusion Models to Data Group}.
\newblock \bibinfo{journal}{\emph{arXiv preprint arXiv:2407.03153}} (\bibinfo{year}{2024}).
\newblock


\bibitem[Lundberg and Lee(2017)]%
        {lundberg2017unified}
\bibfield{author}{\bibinfo{person}{Scott~M Lundberg} {and} \bibinfo{person}{Su-In Lee}.} \bibinfo{year}{2017}\natexlab{}.
\newblock \showarticletitle{A unified approach to interpreting model predictions}.
\newblock \bibinfo{journal}{\emph{Advances in neural information processing systems}}  \bibinfo{volume}{30} (\bibinfo{year}{2017}).
\newblock


\bibitem[Marzi et~al\mbox{.}(2024)]%
        {marzi2024k}
\bibfield{author}{\bibinfo{person}{Giacomo Marzi}, \bibinfo{person}{Marco Balzano}, {and} \bibinfo{person}{Davide Marchiori}.} \bibinfo{year}{2024}\natexlab{}.
\newblock \showarticletitle{K-Alpha Calculator--Krippendorff's Alpha Calculator: A user-friendly tool for computing Krippendorff's Alpha inter-rater reliability coefficient}.
\newblock \bibinfo{journal}{\emph{MethodsX}}  \bibinfo{volume}{12} (\bibinfo{year}{2024}), \bibinfo{pages}{102545}.
\newblock


\bibitem[Mei et~al\mbox{.}(2023)]%
        {mei2023users}
\bibfield{author}{\bibinfo{person}{Alex Mei}, \bibinfo{person}{Michael Saxon}, \bibinfo{person}{Shiyu Chang}, \bibinfo{person}{Zachary~C Lipton}, {and} \bibinfo{person}{William~Yang Wang}.} \bibinfo{year}{2023}\natexlab{}.
\newblock \showarticletitle{Users are the north star for ai transparency}.
\newblock \bibinfo{journal}{\emph{arXiv preprint arXiv:2303.05500}} (\bibinfo{year}{2023}).
\newblock


\bibitem[Miller(2019)]%
        {miller2019explanation}
\bibfield{author}{\bibinfo{person}{Tim Miller}.} \bibinfo{year}{2019}\natexlab{}.
\newblock \showarticletitle{Explanation in artificial intelligence: Insights from the social sciences}.
\newblock \bibinfo{journal}{\emph{Artificial intelligence}}  \bibinfo{volume}{267} (\bibinfo{year}{2019}), \bibinfo{pages}{1--38}.
\newblock


\bibitem[Nauta et~al\mbox{.}(2023)]%
        {nauta2023from}
\bibfield{author}{\bibinfo{person}{Meike Nauta}, \bibinfo{person}{Jan Trienes}, \bibinfo{person}{Shreyasi Pathak}, \bibinfo{person}{Elisa Nguyen}, \bibinfo{person}{Michelle Peters}, \bibinfo{person}{Yasmin Schmitt}, \bibinfo{person}{J\"{o}rg Schl\"{o}tterer}, \bibinfo{person}{Maurice van Keulen}, {and} \bibinfo{person}{Christin Seifert}.} \bibinfo{year}{2023}\natexlab{}.
\newblock \showarticletitle{From Anecdotal Evidence to Quantitative Evaluation Methods: A Systematic Review on Evaluating Explainable AI}.
\newblock \bibinfo{journal}{\emph{ACM Comput. Surv.}} \bibinfo{volume}{55}, \bibinfo{number}{13s}, Article \bibinfo{articleno}{295} (\bibinfo{date}{jul} \bibinfo{year}{2023}), \bibinfo{numpages}{42}~pages.
\newblock
\showISSN{0360-0300}
\href{https://doi.org/10.1145/3583558}{doi:\nolinkurl{10.1145/3583558}}


\bibitem[Naveed et~al\mbox{.}(2023)]%
        {naveed2023comprehensive}
\bibfield{author}{\bibinfo{person}{Humza Naveed}, \bibinfo{person}{Asad~Ullah Khan}, \bibinfo{person}{Shi Qiu}, \bibinfo{person}{Muhammad Saqib}, \bibinfo{person}{Saeed Anwar}, \bibinfo{person}{Muhammad Usman}, \bibinfo{person}{Naveed Akhtar}, \bibinfo{person}{Nick Barnes}, {and} \bibinfo{person}{Ajmal Mian}.} \bibinfo{year}{2023}\natexlab{}.
\newblock \showarticletitle{A comprehensive overview of large language models}.
\newblock \bibinfo{journal}{\emph{arXiv preprint arXiv:2307.06435}} (\bibinfo{year}{2023}).
\newblock


\bibitem[Nguyen et~al\mbox{.}(2023)]%
        {nguyen2023bayesian}
\bibfield{author}{\bibinfo{person}{Elisa Nguyen}, \bibinfo{person}{Minjoon Seo}, {and} \bibinfo{person}{Seong~Joon Oh}.} \bibinfo{year}{2023}\natexlab{}.
\newblock \showarticletitle{A Bayesian Approach To Analysing Training Data Attribution in Deep Learning}. In \bibinfo{booktitle}{\emph{Proceedings of the 2023 Conference on Neural Information Processing Systems}}.
\newblock


\bibitem[OpenAI(2022)]%
        {OpenAI_ChatGPT35}
\bibfield{author}{\bibinfo{person}{OpenAI}.} \bibinfo{year}{2022}\natexlab{}.
\newblock \bibinfo{title}{ChatGPT-3.5}.
\newblock
\urldef\tempurl%
\url{https://chat.openai.com}
\showURL{%
\tempurl}


\bibitem[Park et~al\mbox{.}(2023)]%
        {park2023trak}
\bibfield{author}{\bibinfo{person}{Sung~Min Park}, \bibinfo{person}{Kristian Georgiev}, \bibinfo{person}{Andrew Ilyas}, \bibinfo{person}{Guillaume Leclerc}, {and} \bibinfo{person}{Aleksander Madry}.} \bibinfo{year}{2023}\natexlab{}.
\newblock \showarticletitle{TRAK: Attributing Model Behavior at Scale}. In \bibinfo{booktitle}{\emph{International Conference on Machine Learning (ICML)}}.
\newblock


\bibitem[Parliament(2023)]%
        {aiact}
\bibfield{author}{\bibinfo{person}{European Parliament}.} \bibinfo{year}{2023}\natexlab{}.
\newblock \bibinfo{title}{AI Act, Annex III}.
\newblock


\bibitem[Paszke et~al\mbox{.}(2019)]%
        {pytorch}
\bibfield{author}{\bibinfo{person}{Adam Paszke}, \bibinfo{person}{Sam Gross}, \bibinfo{person}{Francisco Massa}, \bibinfo{person}{Adam Lerer}, \bibinfo{person}{James Bradbury}, \bibinfo{person}{Gregory Chanan}, \bibinfo{person}{Trevor Killeen}, \bibinfo{person}{Zeming Lin}, \bibinfo{person}{Natalia Gimelshein}, \bibinfo{person}{Luca Antiga}, \bibinfo{person}{Alban Desmaison}, \bibinfo{person}{Andreas Kopf}, \bibinfo{person}{Edward Yang}, \bibinfo{person}{Zachary DeVito}, \bibinfo{person}{Martin Raison}, \bibinfo{person}{Alykhan Tejani}, \bibinfo{person}{Sasank Chilamkurthy}, \bibinfo{person}{Benoit Steiner}, \bibinfo{person}{Lu Fang}, \bibinfo{person}{Junjie Bai}, {and} \bibinfo{person}{Soumith Chintala}.} \bibinfo{year}{2019}\natexlab{}.
\newblock \showarticletitle{PyTorch: An Imperative Style, High-Performance Deep Learning Library}.
\newblock In \bibinfo{booktitle}{\emph{Advances in Neural Information Processing Systems 32}}. \bibinfo{publisher}{Curran Associates, Inc.}, \bibinfo{pages}{8024--8035}.
\newblock
\urldef\tempurl%
\url{http://papers.neurips.cc/paper/9015-pytorch-an-imperative-style-high-performance-deep-learning-library.pdf}
\showURL{%
\tempurl}


\bibitem[Pearson(1900)]%
        {pearson1900x}
\bibfield{author}{\bibinfo{person}{Karl Pearson}.} \bibinfo{year}{1900}\natexlab{}.
\newblock \showarticletitle{X. On the criterion that a given system of deviations from the probable in the case of a correlated system of variables is such that it can be reasonably supposed to have arisen from random sampling}.
\newblock \bibinfo{journal}{\emph{The London, Edinburgh, and Dublin Philosophical Magazine and Journal of Science}} \bibinfo{volume}{50}, \bibinfo{number}{302} (\bibinfo{year}{1900}), \bibinfo{pages}{157--175}.
\newblock


\bibitem[Pedregosa et~al\mbox{.}(2011)]%
        {scikit-learn}
\bibfield{author}{\bibinfo{person}{F. Pedregosa}, \bibinfo{person}{G. Varoquaux}, \bibinfo{person}{A. Gramfort}, \bibinfo{person}{V. Michel}, \bibinfo{person}{B. Thirion}, \bibinfo{person}{O. Grisel}, \bibinfo{person}{M. Blondel}, \bibinfo{person}{P. Prettenhofer}, \bibinfo{person}{R. Weiss}, \bibinfo{person}{V. Dubourg}, \bibinfo{person}{J. Vanderplas}, \bibinfo{person}{A. Passos}, \bibinfo{person}{D. Cournapeau}, \bibinfo{person}{M. Brucher}, \bibinfo{person}{M. Perrot}, {and} \bibinfo{person}{E. Duchesnay}.} \bibinfo{year}{2011}\natexlab{}.
\newblock \showarticletitle{Scikit-learn: Machine Learning in {P}ython}.
\newblock \bibinfo{journal}{\emph{Journal of Machine Learning Research}}  \bibinfo{volume}{12} (\bibinfo{year}{2011}), \bibinfo{pages}{2825--2830}.
\newblock


\bibitem[Pezeshkpour et~al\mbox{.}(2022)]%
        {pezeshkpour-etal-2022-combining}
\bibfield{author}{\bibinfo{person}{Pouya Pezeshkpour}, \bibinfo{person}{Sarthak Jain}, \bibinfo{person}{Sameer Singh}, {and} \bibinfo{person}{Byron Wallace}.} \bibinfo{year}{2022}\natexlab{}.
\newblock \showarticletitle{Combining Feature and Instance Attribution to Detect Artifacts}. In \bibinfo{booktitle}{\emph{Findings of the Association for Computational Linguistics: ACL 2022}}. \bibinfo{publisher}{Association for Computational Linguistics}, \bibinfo{address}{Dublin, Ireland}, \bibinfo{pages}{1934--1946}.
\newblock
\href{https://doi.org/10.18653/v1/2022.findings-acl.153}{doi:\nolinkurl{10.18653/v1/2022.findings-acl.153}}


\bibitem[Piorkowski et~al\mbox{.}(2023)]%
        {piorkowski2023aimee}
\bibfield{author}{\bibinfo{person}{David Piorkowski}, \bibinfo{person}{Inge Vejsbjerg}, \bibinfo{person}{Owen Cornec}, \bibinfo{person}{Elizabeth~M. Daly}, {and} \bibinfo{person}{\"{O}znur Alkan}.} \bibinfo{year}{2023}\natexlab{}.
\newblock \showarticletitle{AIMEE: An Exploratory Study of How Rules Support AI Developers to Explain and Edit Models}.
\newblock \bibinfo{journal}{\emph{Proc. ACM Hum.-Comput. Interact.}} \bibinfo{volume}{7}, \bibinfo{number}{CSCW2}, Article \bibinfo{articleno}{255} (\bibinfo{date}{Oct.} \bibinfo{year}{2023}), \bibinfo{numpages}{25}~pages.
\newblock
\href{https://doi.org/10.1145/3610046}{doi:\nolinkurl{10.1145/3610046}}


\bibitem[Pruthi et~al\mbox{.}(2020)]%
        {pruthi_estimating_2020}
\bibfield{author}{\bibinfo{person}{Garima Pruthi}, \bibinfo{person}{Frederick Liu}, \bibinfo{person}{Satyen Kale}, {and} \bibinfo{person}{Mukund Sundararajan}.} \bibinfo{year}{2020}\natexlab{}.
\newblock \showarticletitle{Estimating Training Data Influence by Tracing Gradient Descent}. In \bibinfo{booktitle}{\emph{Advances in Neural Information Processing Systems}}, \bibfield{editor}{\bibinfo{person}{H.~Larochelle}, \bibinfo{person}{M.~Ranzato}, \bibinfo{person}{R.~Hadsell}, \bibinfo{person}{M.F. Balcan}, {and} \bibinfo{person}{H.~Lin}} (Eds.), Vol.~\bibinfo{volume}{33}. \bibinfo{publisher}{Curran Associates, Inc.}, \bibinfo{pages}{19920--19930}.
\newblock
\urldef\tempurl%
\url{https://proceedings.neurips.cc/paper_files/paper/2020/file/e6385d39ec9394f2f3a354d9d2b88eec-Paper.pdf}
\showURL{%
\tempurl}


\bibitem[Radford et~al\mbox{.}(2023)]%
        {whisper}
\bibfield{author}{\bibinfo{person}{Alec Radford}, \bibinfo{person}{Jong~Wook Kim}, \bibinfo{person}{Tao Xu}, \bibinfo{person}{Greg Brockman}, \bibinfo{person}{Christine Mcleavey}, {and} \bibinfo{person}{Ilya Sutskever}.} \bibinfo{year}{2023}\natexlab{}.
\newblock \showarticletitle{Robust Speech Recognition via Large-Scale Weak Supervision}. In \bibinfo{booktitle}{\emph{Proceedings of the 40th International Conference on Machine Learning}} \emph{(\bibinfo{series}{Proceedings of Machine Learning Research}, Vol.~\bibinfo{volume}{202})}, \bibfield{editor}{\bibinfo{person}{Andreas Krause}, \bibinfo{person}{Emma Brunskill}, \bibinfo{person}{Kyunghyun Cho}, \bibinfo{person}{Barbara Engelhardt}, \bibinfo{person}{Sivan Sabato}, {and} \bibinfo{person}{Jonathan Scarlett}} (Eds.). \bibinfo{publisher}{PMLR}, \bibinfo{pages}{28492--28518}.
\newblock
\urldef\tempurl%
\url{https://proceedings.mlr.press/v202/radford23a.html}
\showURL{%
\tempurl}


\bibitem[Ribeiro et~al\mbox{.}(2016)]%
        {ribeiro2016should}
\bibfield{author}{\bibinfo{person}{Marco~Tulio Ribeiro}, \bibinfo{person}{Sameer Singh}, {and} \bibinfo{person}{Carlos Guestrin}.} \bibinfo{year}{2016}\natexlab{}.
\newblock \showarticletitle{" Why should i trust you?" Explaining the predictions of any classifier}. In \bibinfo{booktitle}{\emph{Proceedings of the 22nd ACM SIGKDD international conference on knowledge discovery and data mining}}. \bibinfo{pages}{1135--1144}.
\newblock


\bibitem[Rong et~al\mbox{.}(2023)]%
        {rong2024hcxai}
\bibfield{author}{\bibinfo{person}{Yao Rong}, \bibinfo{person}{Tobias Leemann}, \bibinfo{person}{Thai-Trang Nguyen}, \bibinfo{person}{Lisa Fiedler}, \bibinfo{person}{Peizhu Qian}, \bibinfo{person}{Vaibhav Unhelkar}, \bibinfo{person}{Tina Seidel}, \bibinfo{person}{Gjergji Kasneci}, {and} \bibinfo{person}{Enkelejda Kasneci}.} \bibinfo{year}{2023}\natexlab{}.
\newblock \showarticletitle{Towards Human-Centered Explainable AI: A Survey of User Studies for Model Explanations}.
\newblock \bibinfo{journal}{\emph{IEEE Trans. Pattern Anal. Mach. Intell.}} \bibinfo{volume}{46}, \bibinfo{number}{4} (\bibinfo{date}{nov} \bibinfo{year}{2023}), \bibinfo{pages}{2104–2122}.
\newblock
\showISSN{0162-8828}
\href{https://doi.org/10.1109/TPAMI.2023.3331846}{doi:\nolinkurl{10.1109/TPAMI.2023.3331846}}


\bibitem[Rutjes et~al\mbox{.}(2019)]%
        {rutjes2019considerations}
\bibfield{author}{\bibinfo{person}{Heleen Rutjes}, \bibinfo{person}{Martijn Willemsen}, {and} \bibinfo{person}{Wijnand IJsselsteijn}.} \bibinfo{year}{2019}\natexlab{}.
\newblock \showarticletitle{Considerations on explainable AI and users’ mental models}. In \bibinfo{booktitle}{\emph{CHI 2019 Workshop: Where is the human? Bridging the gap between AI and HCI}}. Association for Computing Machinery, Inc.
\newblock


\bibitem[Sambasivan et~al\mbox{.}(2021)]%
        {sambasivan2021cascades}
\bibfield{author}{\bibinfo{person}{Nithya Sambasivan}, \bibinfo{person}{Shivani Kapania}, \bibinfo{person}{Hannah Highfill}, \bibinfo{person}{Diana Akrong}, \bibinfo{person}{Praveen Paritosh}, {and} \bibinfo{person}{Lora~M Aroyo}.} \bibinfo{year}{2021}\natexlab{}.
\newblock \showarticletitle{“Everyone wants to do the model work, not the data work”: Data Cascades in High-Stakes AI}. In \bibinfo{booktitle}{\emph{Proceedings of the 2021 CHI Conference on Human Factors in Computing Systems}} (<conf-loc>, <city>Yokohama</city>, <country>Japan</country>, </conf-loc>) \emph{(\bibinfo{series}{CHI '21})}. \bibinfo{publisher}{Association for Computing Machinery}, \bibinfo{address}{New York, NY, USA}, Article \bibinfo{articleno}{39}, \bibinfo{numpages}{15}~pages.
\newblock
\showISBNx{9781450380966}
\href{https://doi.org/10.1145/3411764.3445518}{doi:\nolinkurl{10.1145/3411764.3445518}}


\bibitem[Schioppa et~al\mbox{.}(2022)]%
        {schioppa2022arnoldi}
\bibfield{author}{\bibinfo{person}{Andrea Schioppa}, \bibinfo{person}{Polina Zablotskaia}, \bibinfo{person}{David Vilar}, {and} \bibinfo{person}{Artem Sokolov}.} \bibinfo{year}{2022}\natexlab{}.
\newblock \showarticletitle{Scaling Up Influence Functions}.
\newblock \bibinfo{journal}{\emph{Proceedings of the AAAI Conference on Artificial Intelligence}} \bibinfo{volume}{36}, \bibinfo{number}{8} (\bibinfo{date}{Jun.} \bibinfo{year}{2022}), \bibinfo{pages}{8179--8186}.
\newblock
\href{https://doi.org/10.1609/aaai.v36i8.20791}{doi:\nolinkurl{10.1609/aaai.v36i8.20791}}


\bibitem[Shen et~al\mbox{.}(2023)]%
        {shen2023convxai}
\bibfield{author}{\bibinfo{person}{Hua Shen}, \bibinfo{person}{Chieh-Yang Huang}, \bibinfo{person}{Tongshuang Wu}, {and} \bibinfo{person}{Ting-Hao~Kenneth Huang}.} \bibinfo{year}{2023}\natexlab{}.
\newblock \showarticletitle{ConvXAI: Delivering Heterogeneous AI Explanations via Conversations to Support Human-AI Scientific Writing}. In \bibinfo{booktitle}{\emph{Companion Publication of the 2023 Conference on Computer Supported Cooperative Work and Social Computing}} (Minneapolis, MN, USA) \emph{(\bibinfo{series}{CSCW '23 Companion})}. \bibinfo{publisher}{Association for Computing Machinery}, \bibinfo{address}{New York, NY, USA}, \bibinfo{pages}{384–387}.
\newblock
\showISBNx{9798400701290}
\href{https://doi.org/10.1145/3584931.3607492}{doi:\nolinkurl{10.1145/3584931.3607492}}


\bibitem[Simonyan et~al\mbox{.}(2013)]%
        {simonyan2013deep}
\bibfield{author}{\bibinfo{person}{Karen Simonyan}, \bibinfo{person}{Andrea Vedaldi}, {and} \bibinfo{person}{Andrew Zisserman}.} \bibinfo{year}{2013}\natexlab{}.
\newblock \showarticletitle{Deep inside convolutional networks: Visualising image classification models and saliency maps}.
\newblock \bibinfo{journal}{\emph{arXiv preprint arXiv:1312.6034}} (\bibinfo{year}{2013}).
\newblock


\bibitem[Smilkov et~al\mbox{.}(2017)]%
        {smilkov2017smoothgrad}
\bibfield{author}{\bibinfo{person}{Daniel Smilkov}, \bibinfo{person}{Nikhil Thorat}, \bibinfo{person}{Been Kim}, \bibinfo{person}{Fernanda Vi{\'e}gas}, {and} \bibinfo{person}{Martin Wattenberg}.} \bibinfo{year}{2017}\natexlab{}.
\newblock \showarticletitle{Smoothgrad: removing noise by adding noise}.
\newblock \bibinfo{journal}{\emph{arXiv preprint arXiv:1706.03825}} (\bibinfo{year}{2017}).
\newblock


\bibitem[Sundararajan et~al\mbox{.}(2017)]%
        {sundararajan2017axiomatic}
\bibfield{author}{\bibinfo{person}{Mukund Sundararajan}, \bibinfo{person}{Ankur Taly}, {and} \bibinfo{person}{Qiqi Yan}.} \bibinfo{year}{2017}\natexlab{}.
\newblock \showarticletitle{Axiomatic attribution for deep networks}. In \bibinfo{booktitle}{\emph{International conference on machine learning}}. PMLR, \bibinfo{pages}{3319--3328}.
\newblock


\bibitem[Taesiri et~al\mbox{.}(2022)]%
        {taesiri2022visual}
\bibfield{author}{\bibinfo{person}{Mohammad~Reza Taesiri}, \bibinfo{person}{Giang Nguyen}, {and} \bibinfo{person}{Anh Nguyen}.} \bibinfo{year}{2022}\natexlab{}.
\newblock \showarticletitle{Visual correspondence-based explanations improve AI robustness and human-AI team accuracy}.
\newblock \bibinfo{journal}{\emph{Advances in Neural Information Processing Systems}}  \bibinfo{volume}{35} (\bibinfo{year}{2022}), \bibinfo{pages}{34287--34301}.
\newblock


\bibitem[Teso et~al\mbox{.}(2021)]%
        {teso2021interactive}
\bibfield{author}{\bibinfo{person}{Stefano Teso}, \bibinfo{person}{Andrea Bontempelli}, \bibinfo{person}{Fausto Giunchiglia}, {and} \bibinfo{person}{Andrea Passerini}.} \bibinfo{year}{2021}\natexlab{}.
\newblock \showarticletitle{Interactive Label Cleaning with Example-based Explanations}. In \bibinfo{booktitle}{\emph{Advances in Neural Information Processing Systems}}, \bibfield{editor}{\bibinfo{person}{A.~Beygelzimer}, \bibinfo{person}{Y.~Dauphin}, \bibinfo{person}{P.~Liang}, {and} \bibinfo{person}{J.~Wortman Vaughan}} (Eds.).
\newblock
\urldef\tempurl%
\url{https://openreview.net/forum?id=T6m9bNI7C__}
\showURL{%
\tempurl}


\bibitem[Wah et~al\mbox{.}(2011)]%
        {WahCUB_200_2011}
\bibfield{author}{\bibinfo{person}{C. Wah}, \bibinfo{person}{S. Branson}, \bibinfo{person}{P. Welinder}, \bibinfo{person}{P. Perona}, {and} \bibinfo{person}{S. Belongie}.} \bibinfo{year}{2011}\natexlab{}.
\newblock \bibinfo{booktitle}{\emph{The Caltech-UCSD Birds-200-2011 Dataset}}.
\newblock \bibinfo{type}{{T}echnical {R}eport} CNS-TR-2011-001. \bibinfo{institution}{California Institute of Technology}.
\newblock


\bibitem[Wang et~al\mbox{.}(2025a)]%
        {wang2025better}
\bibfield{author}{\bibinfo{person}{Andrew Wang}, \bibinfo{person}{Elisa Nguyen}, \bibinfo{person}{Runshi Yang}, \bibinfo{person}{Juhan Bae}, \bibinfo{person}{Sheila~A McIlraith}, {and} \bibinfo{person}{Roger Grosse}.} \bibinfo{year}{2025}\natexlab{a}.
\newblock \showarticletitle{Better Training Data Attribution via Better Inverse Hessian-Vector Products}.
\newblock \bibinfo{journal}{\emph{arXiv preprint arXiv:2507.14740}} (\bibinfo{year}{2025}).
\newblock


\bibitem[Wang et~al\mbox{.}(2019)]%
        {wang2019designing}
\bibfield{author}{\bibinfo{person}{Danding Wang}, \bibinfo{person}{Qian Yang}, \bibinfo{person}{Ashraf Abdul}, {and} \bibinfo{person}{Brian~Y Lim}.} \bibinfo{year}{2019}\natexlab{}.
\newblock \showarticletitle{Designing theory-driven user-centric explainable AI}. In \bibinfo{booktitle}{\emph{Proceedings of the 2019 CHI conference on human factors in computing systems}}. \bibinfo{pages}{1--15}.
\newblock


\bibitem[Wang et~al\mbox{.}(2024)]%
        {wang2024error}
\bibfield{author}{\bibinfo{person}{Fulton Wang}, \bibinfo{person}{Julius Adebayo}, \bibinfo{person}{Sarah Tan}, \bibinfo{person}{Diego Garcia-Olano}, {and} \bibinfo{person}{Narine Kokhlikyan}.} \bibinfo{year}{2024}\natexlab{}.
\newblock \showarticletitle{Error discovery by clustering influence embeddings}. In \bibinfo{booktitle}{\emph{Proceedings of the 37th International Conference on Neural Information Processing Systems}} (New Orleans, LA, USA) \emph{(\bibinfo{series}{NIPS '23})}. \bibinfo{publisher}{Curran Associates Inc.}, \bibinfo{address}{Red Hook, NY, USA}, Article \bibinfo{articleno}{1809}, \bibinfo{numpages}{13}~pages.
\newblock


\bibitem[Wang et~al\mbox{.}(2025b)]%
        {wang2025capturing}
\bibfield{author}{\bibinfo{person}{Jiachen~T Wang}, \bibinfo{person}{Dawn Song}, \bibinfo{person}{James Zou}, \bibinfo{person}{Prateek Mittal}, {and} \bibinfo{person}{Ruoxi Jia}.} \bibinfo{year}{2025}\natexlab{b}.
\newblock \showarticletitle{Capturing the Temporal Dependence of Training Data Influence}. In \bibinfo{booktitle}{\emph{The Thirteenth International Conference on Learning Representations}}.
\newblock


\bibitem[Weisstein(2004)]%
        {weisstein2004bonferroni}
\bibfield{author}{\bibinfo{person}{Eric~W Weisstein}.} \bibinfo{year}{2004}\natexlab{}.
\newblock \showarticletitle{Bonferroni correction}.
\newblock \bibinfo{journal}{\emph{https://mathworld. wolfram. com/}} (\bibinfo{year}{2004}).
\newblock


\bibitem[Williams(2021)]%
        {williams2021towards}
\bibfield{author}{\bibinfo{person}{Oyindamola Williams}.} \bibinfo{year}{2021}\natexlab{}.
\newblock \showarticletitle{Towards human-centred explainable ai: A systematic literature review}.
\newblock \bibinfo{journal}{\emph{Master’s Thesis}} (\bibinfo{year}{2021}).
\newblock


\bibitem[Wolf(2019)]%
        {wolf2019explainability}
\bibfield{author}{\bibinfo{person}{Christine~T Wolf}.} \bibinfo{year}{2019}\natexlab{}.
\newblock \showarticletitle{Explainability scenarios: towards scenario-based XAI design}. In \bibinfo{booktitle}{\emph{Proceedings of the 24th International Conference on Intelligent User Interfaces}}. \bibinfo{pages}{252--257}.
\newblock


\bibitem[Zheng et~al\mbox{.}(2023)]%
        {zheng2023intriguing}
\bibfield{author}{\bibinfo{person}{Xiaosen Zheng}, \bibinfo{person}{Tianyu Pang}, \bibinfo{person}{Chao Du}, \bibinfo{person}{Jing Jiang}, {and} \bibinfo{person}{Min Lin}.} \bibinfo{year}{2023}\natexlab{}.
\newblock \showarticletitle{Intriguing properties of data attribution on diffusion models}.
\newblock \bibinfo{journal}{\emph{arXiv preprint arXiv:2311.00500}} (\bibinfo{year}{2023}).
\newblock


\end{thebibliography}
